%% file: ms.tex
\documentclass[12pt,preprint]{aastex}
\begin{document}
\newcommand{\msun}{\mbox{M$_{\odot}$}}
\newcommand{\rsun}{\mbox{R$_{\odot}$}}
\newcommand{\zsun}{\mbox{Z$_{\odot}$}}
\newcommand{\lsun}{\mbox{L$_{\odot}$}} 

\title{Towards an Accurate Determination of Parameters for Very Massive
Stars: the Eclipsing Binary LMC-SC1-105\altaffilmark{1}}

\author{Alceste Z. Bonanos\altaffilmark{2}}

\affil{Carnegie Institution of Washington, Department
of Terrestrial Magnetism, \\ 5241 Broad Branch Rd. NW, Washington D.C.,
20015, USA \\ \tt e-mail: bonanos@dtm.ciw.edu}

\altaffiltext{1}{Based on observations obtained with the 2.5 meter
DuPont and 6.5 meter Magellan Clay Telescope located at Las Campanas
Observatory, Chile.}

\altaffiltext{2}{Vera Rubin Fellow}

\begin{abstract}

This paper presents a photometric and spectroscopic study of the bright
blue eclipsing binary LMC-SC1-105, selected from the OGLE catalog as a
candidate host of very massive stars ($\geq 30\,\msun$). The system is
found to be a double-lined spectroscopic binary, which indeed contains
massive stars. The masses and radii of the components are $\rm
M_{1}=30.9\pm1.0\;\msun$, $\rm M_{2}=13.0\pm0.7\;\msun$, and $\rm
R_{1}=15.1\pm0.2\;\rsun$, $\rm R_{2}=11.9\pm0.2\;\rsun$, respectively.
The less massive star is found to be filling its Roche lobe, indicating
the system has undergone mass-transfer. The spectra of LMC-SC1-105
display the Struve-Sahade effect, with the \ion{He}{1} lines of the
secondary appearing stronger when it is receding and causing the
spectral types to change with phase (O8$+$O8 to O7$+$O8.5). This effect
could be related to the mass-transfer in this system. To date, accurate
($\leq 10\%$) fundamental parameters have only been measured for 15
stars with masses greater than 30~\msun, with the reported measurements
contributing valuable data on the fundamental parameters of very massive
stars at low metallicity. The results of this work demonstrate that the
strategy of targeting the brightest blue stars in eclipsing binaries is
an effective way of studying very massive stars.

\end{abstract}

\keywords{binaries: eclipsing -- binaries: spectroscopic -- stars:
fundamental parameters, individual (OGLE053448.26-694236.4) -- galaxies:
individual (LMC)}

\section{Introduction}
\label{section:intro}
\vspace{-0.4cm}
The fundamental parameters of very massive stars ($\geq 30\,\msun$)
remain uncertain, despite the large impact massive stars have in
astrophysics, both individually and collectively \citep[see review
by][]{Massey03}. The equations of stellar structure allow for stars with
arbitrarily large masses, however the mechanisms to form massive stars
\citep[accretion and mergers; e.g.][]{Bally05} and the associated
instabilities \citep[see][and references therein]{Elmegreen00,
Zinnecker07} are not well understood, hindering theoretical predictions
on the existence of an upper limit on the stellar mass. The ``mass
discrepancy'' problem, i.e. the disagreement between masses derived from
parameters determined by fitting stellar atmosphere models to spectra
and from evolutionary tracks \citep[see e.g.][for a
comparison]{Repolust04, Massey05}, still affects studies of single
massive stars, even though significant progress has been made in both
stellar atmosphere \citep[see review by][]{Herrero07} and stellar
evolution models \citep[e.g.][]{Meynet03}. The parameters of single
stars also suffer from suspected multiplicity, which in many cases
cannot be determined.  Pismis 24-1 demonstrates this problem: its
inferred evolutionary mass $>200\,\msun$ \citep{Walborn02} contradicted
the upper stellar mass limit of $\sim 150\,\msun$ suggested by
statistical arguments based on observations
\citep{Figer05,Oey05}. \citet{Maiz-Apellaniz07} resolved it into a
visual binary with the {\it Hubble Space Telescope}, thereby removing
the discrepancy. One of its components is also a spectroscopic binary,
illustrating the systematic effects often accompanying ``single'' stars.

\vspace{-0.2cm}
The only model-independent way to obtain accurate fundamental parameters
of distant massive stars and to resolve the ``mass discrepancy'' problem
is to use eclipsing binaries \citep[see review by][]{Andersen91}. In
particular, double-lined spectroscopic binary systems exhibiting
eclipses in their light curves are extremely powerful tools for
measuring masses and radii of stars. Specifically, the light curve
provides the orbital period, inclination, eccentricity, the fractional
radii and flux ratio of the two stars. The radial velocity
semi-amplitudes determine the mass ratio; the individual masses can be
solved for by using Kepler's third law. Furthermore, by fitting
synthetic spectra to the observed ones, one can infer the effective
temperatures of the stars, solve for their luminosities and derive the
distance \citep[e.g.\@][]{Bonanos06}. The most massive stars measured in
eclipsing binaries are galactic Wolf-Rayet stars of WN6ha spectral type:
NGC3603-A1 \citep[M$_1=116 \pm 31\; \msun$, M$_2=89 \pm 16\;
\msun$;][]{Schnurr08}, and WR~20a \citep[M$_1=83.0 \pm 5.0\; \msun$ and
M$_2=82.0 \pm 5.0\; \msun$][]{Rauw04, Bonanos04} in Westerlund~2,
presenting a challenge for both stellar evolution and massive star
formation models \citep{Yungelson08, Zinnecker07} and raising the issue
of the frequency and origin of ``binary twins'' \citep{Pinsonneault06,
Lucy06, Krumholz07}. Such systems are of particular interest, since
massive binaries might be progenitors of gamma-ray bursts
\citep[e.g.\@][]{Fryer07}, especially in the case of Population III,
metal-free stars \citep[see][]{Bromm06}.

Analogs of these heavyweight champions, if not more massive binaries,
are bound to exist in the young massive clusters at the center of the
Galaxy (Center, Arches, Quintuplet), in nearby super star clusters
(e.g.\@ Westerlund\,1, R136), in Local Group galaxies (e.g.\@ LMC, SMC,
M31, M33) and beyond (e.g.\@ M81, M83, NGC 2403). A systematic
wide-ranging survey of these clusters and galaxies is currently
underway. The goal is to provide data with which to test star formation
theories, stellar atmosphere and stellar evolution models for both
single and binary stars as a function of metallicity, and the
theoretical predictions on the upper limit of the stellar mass. The
adopted strategy involves two steps: a variability survey to discover
eclipsing binaries in these massive clusters and nearby galaxies, which
is followed by spectroscopy to derive parameters of the brightest --
thus most luminous and massive -- blue systems. However, characterizing
massive stars requires the availability of 8-m class telescopes and high
resolution near-infrared spectrographs (since massive stars in the
Galaxy are extincted and extragalactic ones are faint) and has only
become feasible in the past few years. \citet{Bonanos07} demonstrated
that this method efficiently finds massive candidates, by performing the
first variability survey of the Westerlund~1 super star cluster and
discovering 4 massive eclipsing binary systems.

\begin{figure}[ht]  
\plotone{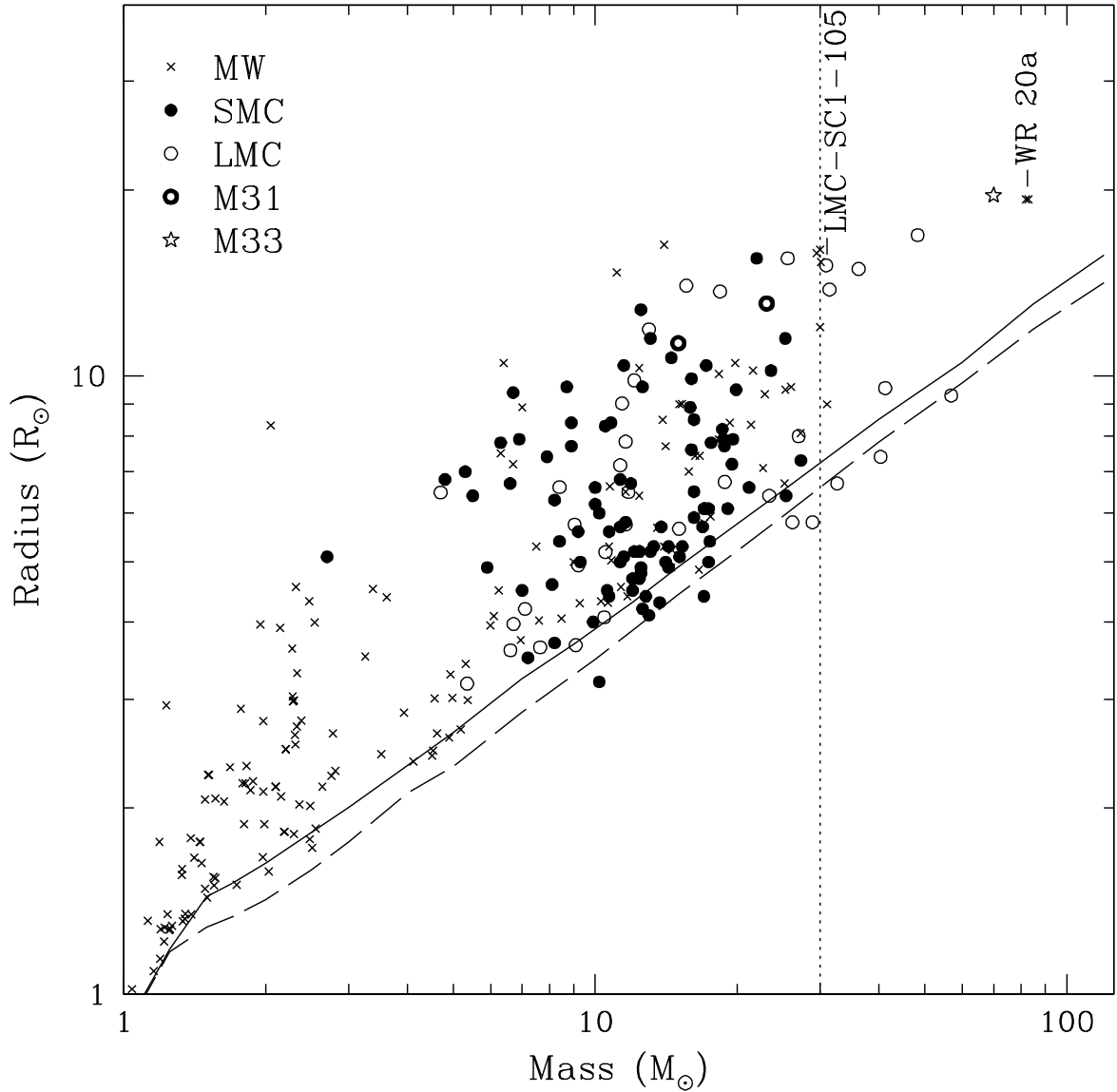}
\caption{Mass and radius determinations of stars in eclipsing binaries,
accurate to $\leq 10\%$ and complete $\geq 30\,\msun$ from the
literature (see \S\ref{section:intro} and Table~\ref{massradiusdata} for
references). The solid line is the Z=0.02 ZAMS from \citet{Schaller92};
the dashed line is the Z=0.008 ZAMS from \citet{Schaerer93}. Note the
small number of measurements for stars with masses greater than
$30\,\msun$, all published since 2001.}
\label{massradius}
\end{figure}

Figure~\ref{massradius} illustrates the extent of our knowledge of
precise fundamental parameters of massive stars. It presents published
mass-radius measurements from eclipsing binaries, accurate to better
than $10\%$ for the more massive component. The zero-age main sequences
(ZAMS) at both Z=0.02 \citep{Schaller92} and Z=0.008 \citep{Schaerer93}
are overplotted as a reference. The Galactic data are mainly taken from
the compilations of \citet{Andersen91} and \citet{Gies03} with additions
and updates from \citet{Vitrichenko07}, but also \citet{Gonzalez05} for
the Large Magellanic Cloud (LMC), \citet{Harries03} and
\citet{Hilditch05} for the SMC, and \citet{Ribas05} for M31. A
literature search was done to include all accurate measurements of stars
in eclipsing binaries with masses $\geq 30\,\msun$, which are presented
in Table~\ref{massradiusdata}. This Table is, to my knowledge, complete
at present and consists of only 14 very massive stars with better than
$10\%$ mass-radius measurements, located in 3 galaxies. Of these, WR~20a
and M33 X-7 \citep{Orosz07} are the most massive and noteworthy. M33 X-7
contains a very massive $70.0 \pm 6.9\; \msun$ O-type giant and a
record-breaking $15.65\;\msun$ black hole, challenging current
evolutionary models, which fail to explain such a large black hole
mass. Without accurate measurements for a large sample of massive stars,
theoretical models will remain unconstrained.

A survey to determine accurate parameters for several massive eclipsing
binaries in the low metallicity (Z$=0.008$) LMC was undertaken, with the
purpose of increasing the sample and improving our understanding of
these rare systems. Several candidates were selected from the OGLE-II
catalog of eclipsing binaries in the LMC \citep{Wyrzykowski03} as the
brightest systems with $B-V<0$. LMC-SC1-105, or OGLE053448.26-694236.4,
is one of the brightest eclipsing binaries with $I_{max}=13.04$~mag,
$V_{max}=12.97$~mag, $B_{max}=12.81$~mag and a preliminary semi-detached
classification. This work presents the analysis of the follow-up
observations obtained for LMC-SC1-105. The paper is organized as
follows: \S2 describes the spectroscopy and data reduction, \S3 the
spectral classification, \S4 the radial velocity curve, \S5 the light
curve analysis, \S6 the evolutionary status, and \S7 the conclusion.

\section{Spectroscopy}
\vspace{-0.2cm}

A total of 9 spectra of LMC-SC1-105 near quadrature phases were acquired
over 4 runs on 2 telescopes at Las Campanas Observatory, Chile. In
December 2005, January and February 2006, spectra were obtained with the
Echelle spectrograph on the 2.5-m DuPont telescope. The $1\arcsec \times
4\arcsec$ slit resulted in a spectral resolution of 10 km~s$^{-1}$
($R=30000$) or 0.15\AA\, at 4500\AA, as measured from the full width
half maximum of the comparison lamp lines. In January 2006 and October
2007, additional spectra were obtained with the blue and red sides of
the MIKE spectrograph \citep{Bernstein03} on the 6.5-m Magellan Clay
telescope. The $1.0\arcsec \times 5.0\arcsec$ slit used in 2006 resulted
in a spectral resolution of 9~km~s$^{-1}$ ($R=32000$) or 0.14\AA\, at
4500\AA, as measured from the full width half maximum of the comparison
lamp lines. The $0.7\arcsec \times 5.0\arcsec$ slit used in 2007
resulted in a spectral resolution of 7~km~s$^{-1}$ ($R=41000$) or
0.11\AA\, at 4500\AA. Table~\ref{speclog} summarizes the log of the
observations, specifying the telescope and instrument used, the exposure
times and final signal to noise (S/N) ratio per pixel measured on the
merged spectra.

The Magellan spectra were extracted using the MIKE reduction pipeline
written by D. Kelson \citep{Kelson00,Kelson03}. The extracted orders for
each star were averaged, normalized and merged.  The wavelength coverage
of the final merged spectra is $3900-5050$\AA\, in the blue and
$5040-7150$\AA\, in the red. The DuPont spectra were reduced, extracted
and wavelength calibrated using the $noao.imred.echelle$ package in
IRAF\footnote{IRAF is distributed by the NOAO, which are operated by the
Association of Universities for Research in Astronomy, Inc., under
cooperative agreement with the NSF.}. Cosmic rays were removed from the
two dimensional images with the algorithm of \citet{Pych04}. The orders
were finally averaged, normalized and merged, yielding a wavelength
range $3700-9000$\AA. The heliocentric radial velocity corrections for
each star were computed with the IRAF $rvsao.bcvcorr$ routine and taken
into account in the subsequent radial velocity determination. Narrow
nebular emission, likely originating from a nearby HII region, is
present in the spectra and was removed. Finally, the \ion{Ca}{2} H and K
and \ion{Na}{1} D lines exhibit multiple absorption features,
corresponding to Galactic and LMC interstellar material with radial
velocities between 0 and 315 km~s$^{-1}$.

\section{Struve-Sahade Effect and Spectral Classification}
\label{section:sse}

\begin{figure}[ht]  
\plotone{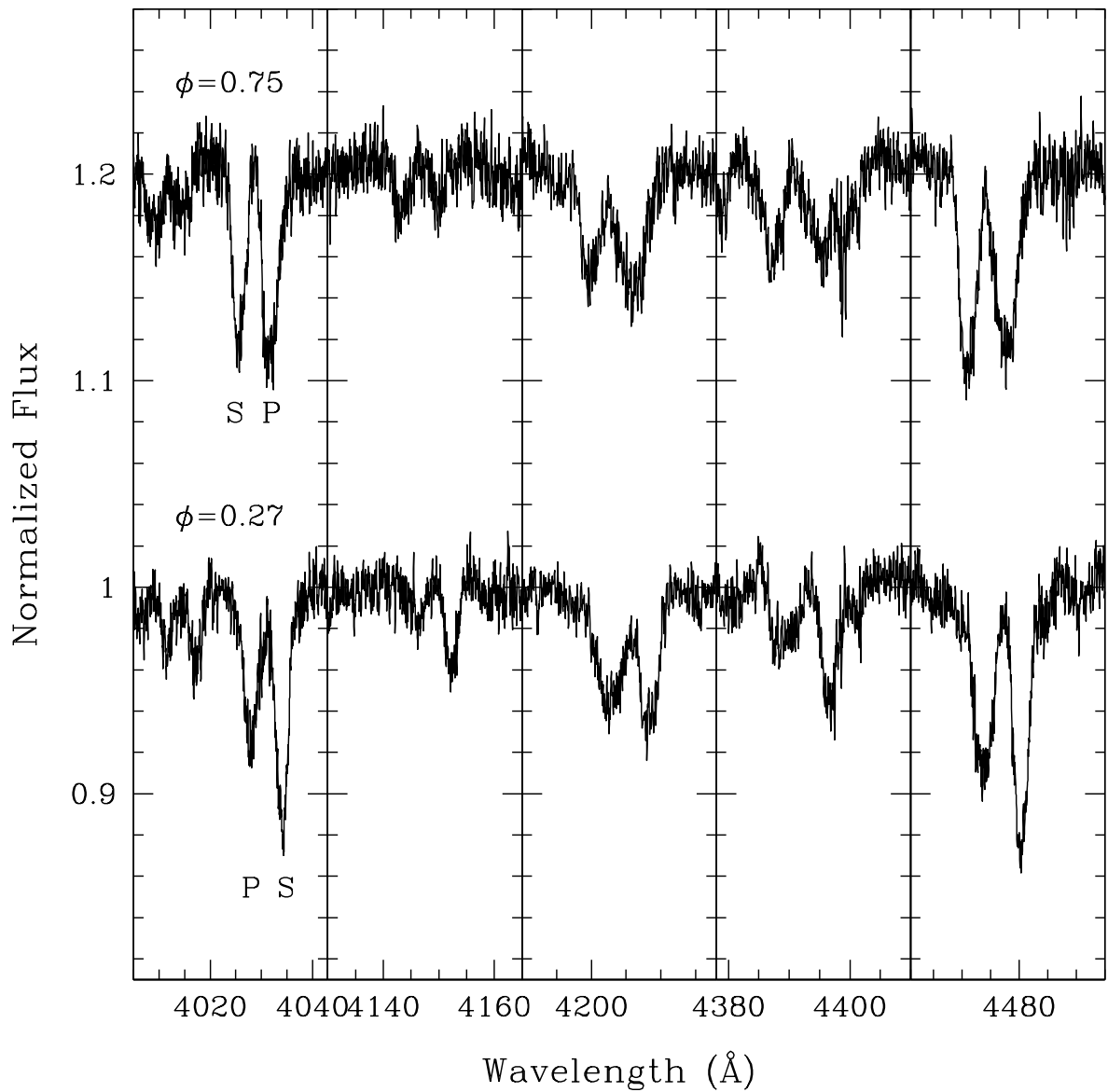}
\caption{Quadrature spectra of LMC-SC1-105, obtained with MIKE/Magellan
at phases $\phi=0.75$ and 0.27, displaying the Struve-Sahade effect. The
primary (P) and secondary (S) stars are labelled in the first panel. The
panels display the \ion{He}{1} $\lambda4009$ and \ion{He}{2}
$\lambda4026$, \ion{He}{1} $\lambda4144$, \ion{He}{2} $\lambda4200$,
\ion{He}{1} $\lambda4387$, \ion{He}{1} $\lambda4471$ lines,
respectively. Note that the lines of the secondary are all stronger at
$\phi=0.27$; the systemic velocity is 284 km~s$^{-1}$; tickmarks
correspond to 5\,\AA\, in the last panel.}
\label{lmc105ss}
\end{figure}

\begin{figure}[ht]  
\plotone{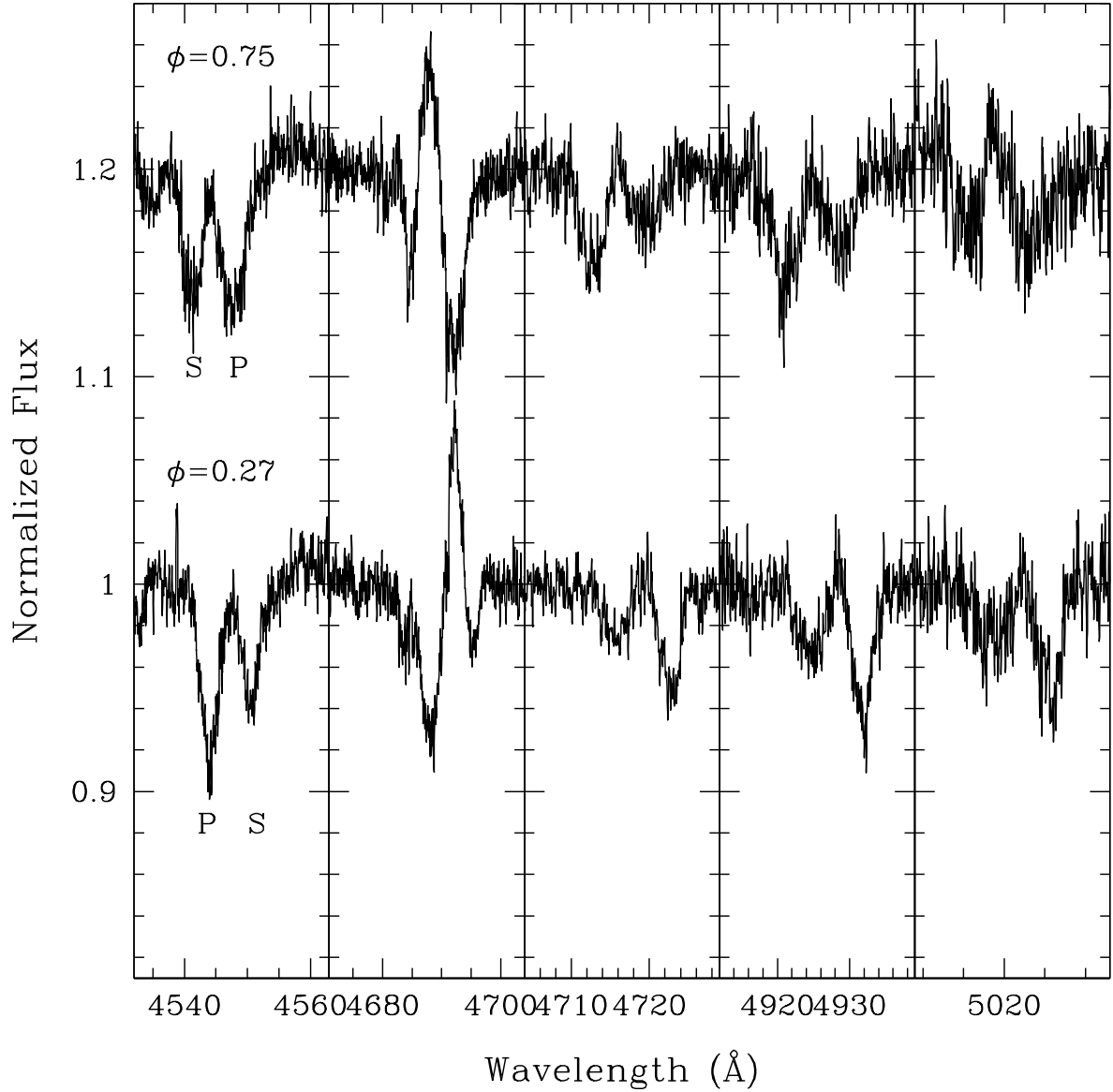}
\caption{The Struve-Sahade effect in LMC-SC1-105 quadrature spectra, as
in Figure~\ref{lmc105ss}, but for the following lines: \ion{He}{2}
$\lambda4541$, \ion{He}{2} $\lambda4686$, \ion{He}{1} $\lambda4713$,
\ion{He}{1} $\lambda4922$, \ion{He}{1} $\lambda5016$, respectively. The
\ion{He}{2} $\lambda4686$ emission is likely due to gas being transfered
from the secondary onto the primary; tickmarks correspond to 5\,\AA\, in
the last panel.}
\label{lmc105ss2}
\end{figure}

\begin{figure}[ht]  
\plotone{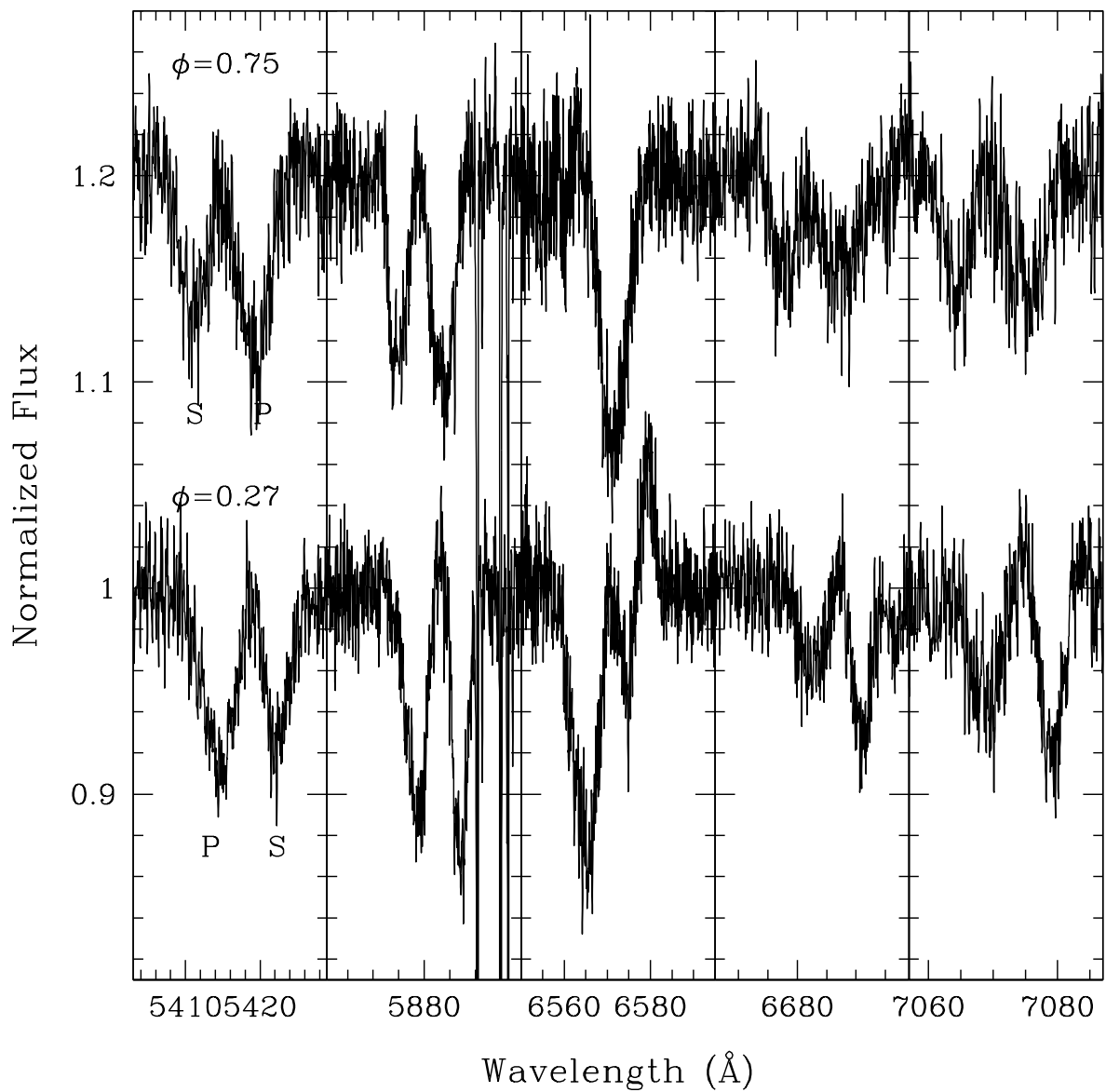}
\caption{The Struve-Sahade effect in LMC-SC1-105 quadrature spectra, as
in Figure~\ref{lmc105ss}, but for the following lines: \ion{He}{2}
$\lambda5411$, \ion{He}{1} $\lambda5876$, H$\alpha$ $\lambda6563$,
\ion{He}{1} $\lambda6678$, \ion{He}{1} $\lambda7065$, respectively. Note
the narrow Galactic and LMC interstellar \ion{Na}{1} D lines at
$\lambda\lambda5890-95$. The S/N of the spectra decreases at redder
wavelengths, due to the lower sensitivity of MIKE and the color of the
system. Tickmarks correspond to 5\,\AA\, in the second and fourth
panels.}
\label{lmc105ss3}
\end{figure}

Inspection of the quadrature spectra reveals that LMC-SC1-105 exhibits
the {\it ``Struve-Sahade effect''} \citep{Stickland97}. This term
describes the variable strength of the spectral lines of the secondary
star (or primary star in some cases) in a double-lined spectroscopic
binary \citep[see][and references
therein]{Howarth97}. Figures~\ref{lmc105ss}, \ref{lmc105ss2}, and
\ref{lmc105ss3} present quadrature spectra of LMC-SC1-105 for the most
prominent \ion{He}{1} and \ion{He}{2} lines. At phase 0.27, the
\ion{He}{1} lines of the secondary are significantly stronger than at
phase 0.75, while, of the \ion{He}{2} lines this is the case only for
\ion{He}{2} $\lambda4200$. Several mechanisms have been proposed to
cause this effect \citep[see][and references therein]{Bagnuolo99,
Linder07}, which could, in principle, affect the luminosity ratio and
masses derived. In the case of LMC-SC1-105, both lines are clearly
resolved and therefore the position of the line centers, and
consequently the radial velocities can be unambiguously measured.
Studying the Struve-Sahade effect in eclipsing binaries could be
valuable for understanding its origin, because the inclination and sizes
of stars are additionally known. A thorough investigation of the effect,
as undertaken by \citet{Linder07}, is beyond the scope of this paper.

\begin{figure}[ht]  
\includegraphics[angle=180,width=6.5in]{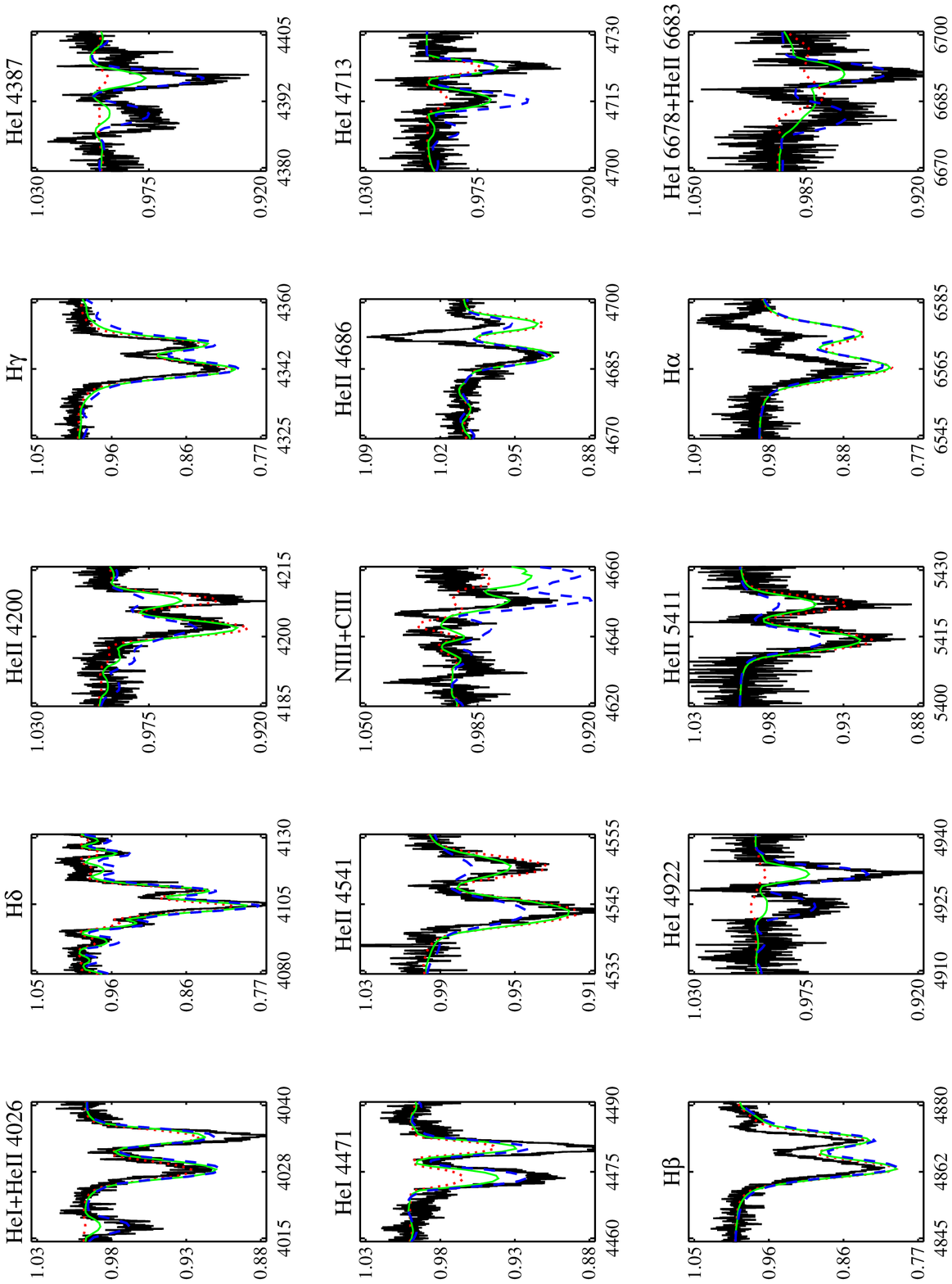}
\caption{{\sc TLUSTY} model compared with the MIKE spectrum of
LMC-SC1-105 at phase 0.27. The colored lines represent the sum of two
models shifted at the appropriate velocity, with the following (T$_{\rm
eff1}$,T$_{\rm eff2}$) pairs (in kK): (37.5, 35) in red (dotted line),
(35, 32.5) in green (solid line), and (32.5, 30) in blue (dashed
line). The green model fits most lines best.}
\label{fitmultioct23}
\end{figure}

\begin{figure}[ht]  
\includegraphics[angle=180,width=6.5in]{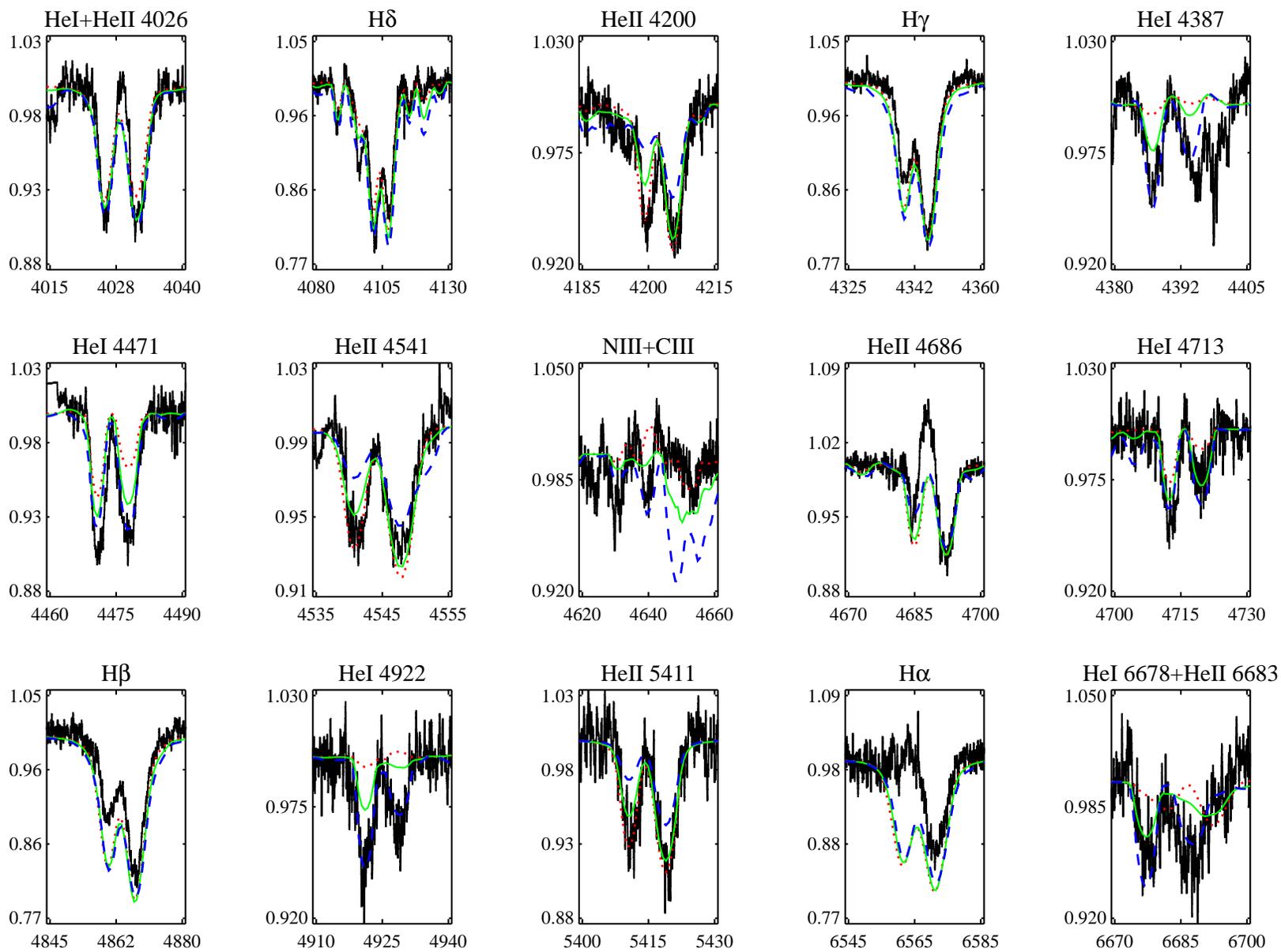}
\caption{Same as Figure~\ref{fitmultioct23}, but for phase 0.75.}
\label{fitmultioct25}
\end{figure}

Following the criteria of \citet{Walborn90}, the spectral types of the
primary and secondary are O8V and O8III-V at phase 0.75,
respectively. The luminosity class of the secondary cannot be
unambiguously determined, because of the strong emission in the
\ion{He}{2} $\lambda4686$ line, superposed on the absorption line of the
secondary. The Struve-Sahade effect further causes the spectral types of
both stars to change. At phase 0.25 the stars appear to have types O7V
and O8.5III-V.  \citet{Massey00} have assigned a O8.5III spectral type
to LMC-SC1-105 (or W28-22, LH 81-72) from their lower resolution
spectra, likely having observed it near or during primary
eclipse. According to the calibration of effective temperature (T$_{\rm
eff}$) with spectral type by \citet{Mokiem07} for the LMC, the primary
should have T$_{\rm eff}=35-40$kK and the secondary T$_{\rm
eff}=30-35$kK. A (sub)giant classification would make it $\sim1-2$kK
cooler than a dwarf with the same spectral class \citep[see
e.g.][]{Martins05}. Figures~\ref{fitmultioct23} and~\ref{fitmultioct25}
present the hydrogen and helium lines at each quadrature and {\sc
TLUSTY} model atmospheres (see \S\ref{section:rv}) for a range of
T$_{\rm eff}$. The combination that best fits the data has T$_{\rm
eff1}=35\pm2.5$kK and T$_{\rm eff2}=32.5\pm2.5$kK, with $\log(g)$ fixed
to 3.50 and the projected rotational velocities $vsini$ to the
synchronous values of 180 km~s$^{-1}$ and 140 km~s$^{-1}$ (as determined
in \S\ref{section:lc}). These T$_{\rm eff}$ values are consistent with
the spectral type calibration of \citet{Mokiem07}. The T$_{\rm eff}$
grid step was adopted as a conservative error. Note that no combination
of models can fit all the helium lines satisfactorily, in particular the
singlet lines \ion{He}{1} $\lambda4387$ and $\lambda4922$.
\citet{Puls05} point out that CMFGEN \citep{Hillier98}, and therefore
{\sc TLUSTY} \citep[since both models are consistent, see][]{Bouret03},
predicts much weaker \ion{He}{1} singlet lines than FASTWIND
\citep{Santolaya97, Puls05}. \citet{Bouret03} also state that ``a
simultaneous fit to all \ion{He}{1} and \ion{He}{2} lines is almost
never achieved''. Computing a finer grid of {\sc TLUSTY} models to
better constrain the T$_{\rm eff}$ of the stars or using a unified model
atmosphere, such as FASTWIND, was not pursued, because the determination
of masses and radii is independent of the T$_{\rm eff}$.

\section{Radial Velocity Curve}
\label{section:rv}

Two methods were used to measure the orbital parameters of LMC-SC1-105:
two dimensional cross correlation (or TODCOR) and spectral
disentangling. TODCOR was developed by \citet{Zucker94} and can
distinguish small velocity separations even more accurately than one
dimensional cross correlation \citep{Tonry79}. Synthetic spectra from
the OSTAR2002 {\sc TLUSTY} non-LTE grid \citep{Lanz03} at half-solar
metallicity were used for the cross correlation. The microturbulent
velocity was fixed to 10 km~s$^{-1}$ and the helium abundance to the
solar value, $He/H= 0.1$ by number. The T$_{\rm eff}$ of the models in
the grid range from 27500--55000 K in steps of 2500 K. The surface
gravity $\log(g)$, depending on the exact T$_{\rm eff}$ value, ranges
from 3.00--4.75 dex (cgs) in steps of 0.25 dex and the microturbulent
velocity was fixed at 10 km~s$^{-1}$. The models were rotationally
broadened (20-400 km~s$^{-1}$ in steps of 20 km~s$^{-1}$) and the
instrumental broadening was applied with the $rotin3$ program
distributed with the {\sc TLUSTY} grid.

Initially, best fit models were computed by minimizing the residuals of
the sum of 2 models shifted appropriately to the highest S/N quadrature
spectrum from UT 2007 October 24. This procedure yielded the following
best fit models (T$_{\rm eff}$, $\log(g)$, $vsini$): (35000, 3.50, 160)
for the primary and (30000, 3.00, 140) for the secondary. These models
were used as templates in TODCOR to derive initial values for the radial
velocities, orbital elements and stellar parameters. Subsequently,
$vsini$ and $\log(g)$ were fixed for the estimation of T$_{\rm
eff}$. The models used for the final analysis are shown in
Figures~\ref{fitmultioct23} and~\ref{fitmultioct25}. The regions around
the Balmer lines and $\lambda\lambda4620-4700$ were excluded from the
TODCOR analysis, as the former are broad and the latter region contains
\ion{He}{2} $\lambda$4686 emission (formed in the wind or interaction
region), which the hydrostatic equilibrium {\sc TLUSTY} models cannot
reproduce. The resulting TODCOR velocities are given in
Table~\ref{rv}. 

An orbital fit for the systemic $\gamma$ velocity, the semi-major axis
and the mass ratio $q$ was performed with PHOEBE \citep[version
0.31a,][]{Prsa05}, which builds on and enhances the capabilities of the
Wilson-Devinney program \citep{Wilson71, Wilson79, Wilson90}. The two
highest S/N spectra were assigned a weight of 2, while the lowest S/N
spectrum was assigned a weight of 0.5; the rest were assigned weights of
1. The values and their formal uncertainties, found by fixing the
ephemeris from the OGLE catalog, are: $q=0.42\pm0.02$, $a \sin
i=38.9\pm0.5\,\rsun$, $\gamma=284\pm3\, \rm km\; s^{-1}$. These imply
semi-amplitudes of $K_{1}=137\pm4$ km~s$^{-1}$ and $K_{2}=326\pm3$
km~s$^{-1}$, and minimum masses of $M_{1} \sin^3 (i)=30.9\pm1.0\,\msun$
and $M_{2} \sin^3 (i)=13.0\pm0.7\,\msun$. The rms of the fit is 12~$\rm
km\; s^{-1}$ for the primary and 8~$\rm km\; s^{-1}$ for the secondary,
which are adopted as representative values for the error in each radial
velocity measurement. Separate $\gamma$ velocities for each star were
also fit for, but yielded values consistent within errors with the value
above, therefore were not considered further.

Another accurate method for deriving radial velocities of spectroscopic
binary stars is spectral disentangling \citep{Simon94}. The program
KOREL \citep{Hadrava95} implements the method using Fourier
transforms. In order to explore the $\chi^2$ parameter space, KOREL was
run for a range of primary semi-amplitudes ($100-160$ km~s$^{-1}$) and
mass ratios ($q=M_{2}/M_{1}: 0.2-0.7$), using the range
$\lambda\lambda3990-4965$, a 4 km~s$^{-1}$ velocity step, excluding the
$\lambda\lambda4620-4700$ region. The resulting values were found to be
$K_1=140^{+15}_{-20}$ km~s$^{-1}$, $q=0.45^{+0.06}_{-0.08}$, in
agreement with the TODCOR results. The latter have more realistic errors
and were adopted in the subsequent analysis. \citet{Southworth07}
present the first detailed comparison of the methods developed to derive
radial velocities and find that disentangling is the most accurate,
however they applied TODCOR on a single line.

\section{Light Curve Analysis}
\label{section:lc}

Besides the OGLE $I-$band light curve, blue and red filter light curves
for LMC-SC1-105, roughly corresponding to $V$ and $R$, are also
available from the MACHO database (ID 81.8881.21). The MACHO light
curves have 331 measurements in the red and 167 in the blue and span 7.2
years, but are noisier than the OGLE-II and OGLE-III light curve
\citep[kindly made available by I. Soszynski; see][]{Soszynski08}, which
has 750 points (after removing outliers) that span more than 11
years. The instrumental OGLE-III light curve was offset to match the
OGLE-II light curve in the analysis below.

Detailed simultaneous modeling of the 3 light curves was performed with
PHOEBE. Both the detached mode and the semi-detached mode with the
secondary filling its Roche lobe were considered. Note, the primary star
(star 1) in both binaries analyzed herein is defined photometrically, as
the hotter star producing the deeper eclipse at phase zero. The
parameters that were allowed to vary are: the inclination $i$, T$_{\rm
eff2}$, the period $P$, and the surface potentials $\Omega_1$ and
$\Omega_2$. The time of primary eclipse T$_0$ was fixed to the value
determined by \citet{Wyrzykowski03}, while the value of the mass ratio
and semi-major axis were fixed to the values determined from the orbital
fit. The light ratio was computed rather than fit for, following
\citet{Prsa05}. Synchronous rotation, a circular orbit and no third
light were assumed, as there was no evidence to the contrary. T$_{\rm
eff1}$ was fixed to the value determined from the spectra in
\S\ref{section:sse}, while the gravity darkening exponents were set to 1
for stars with radiative envelopes (i.e. $T_{\rm eff} \propto g^{0.25}$)
and the albedo values to 0.5, following \citet{Hilditch01}. Limb
darkening coefficients for the square root law were fixed from
\citet{Claret00} and the approximate reflection model was used.

\begin{figure}[ht]  
\plotone{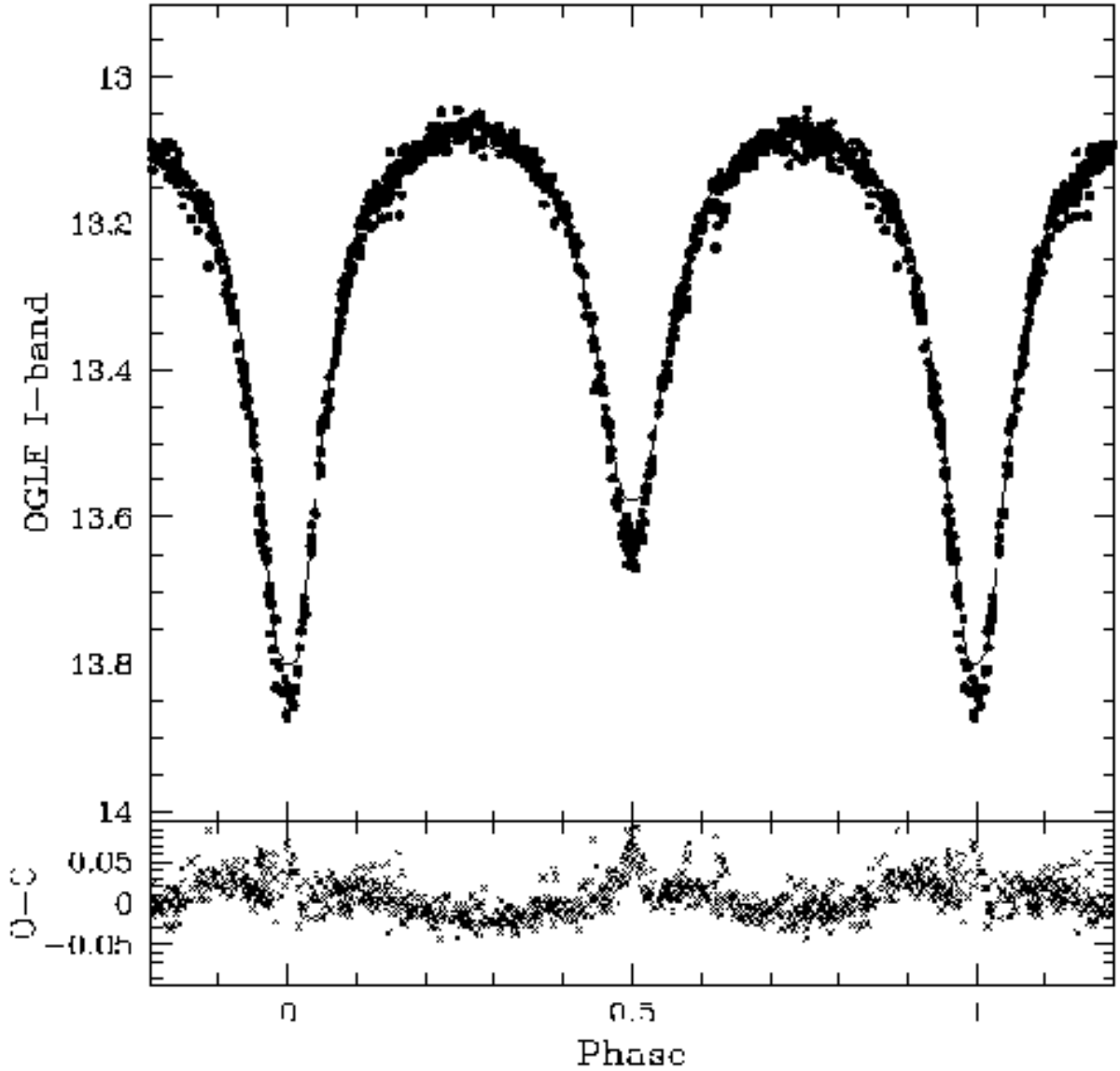}
\caption{Phased OGLE $I-$band light curve of LMC-SC1-105. The best fit
model from PHOEBE (solid curve) assumes a semi-detached configuration
with the secondary filling its Roche lobe. The residuals suggest the
presence of an accretion stream and hot spots (not modeled), arising
from mass-transfer onto the primary.}
\label{lmc105i}
\end{figure}

\begin{figure}[ht]  
\plotone{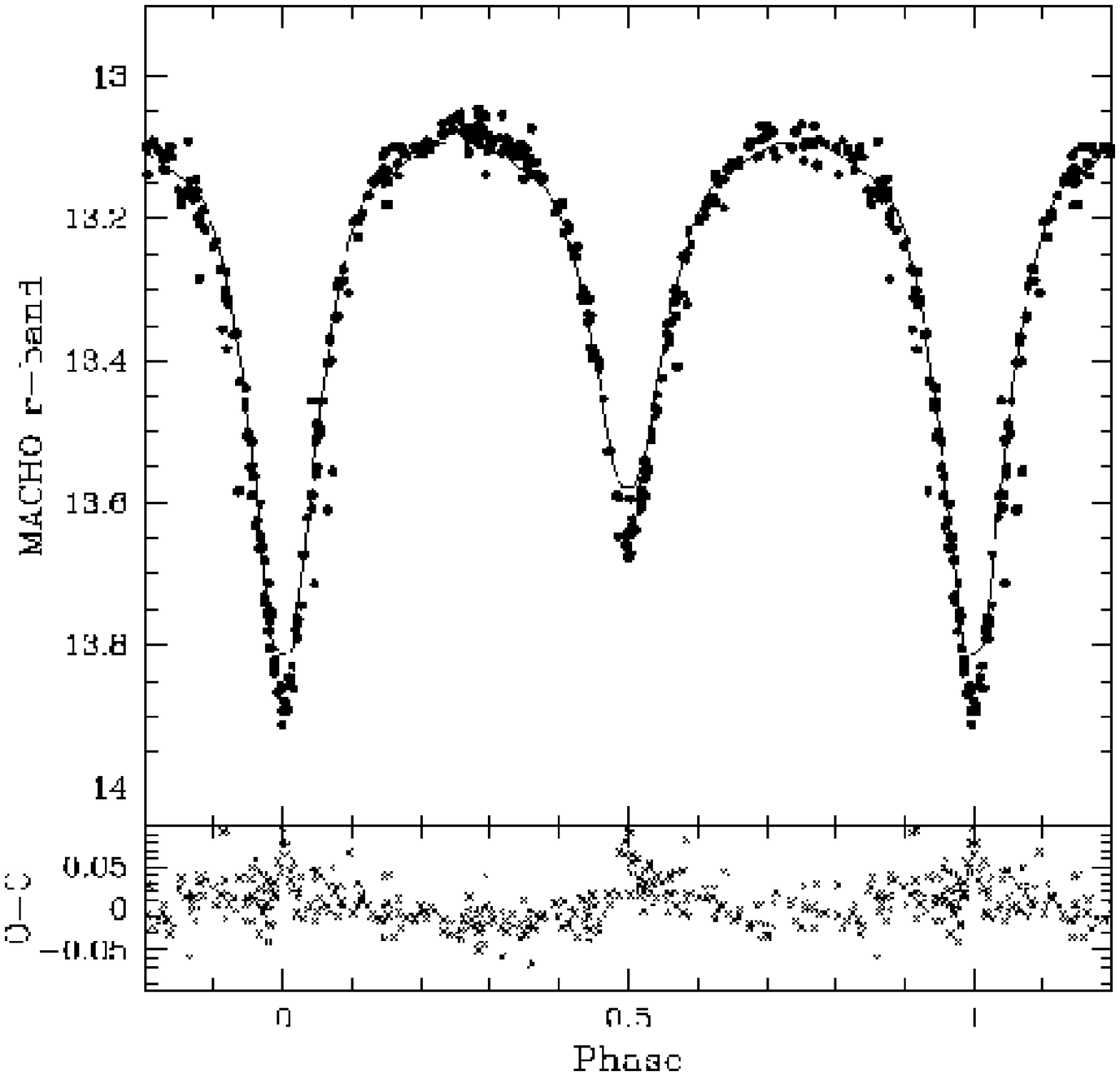}
\caption{Phased MACHO $r$ light curve of LMC-SC1-105 offset to match
OGLE $I-$band photometry; the best fit model from PHOEBE (solid curve)
is overplotted.}
\label{phoebemachorLC}
\end{figure}

\begin{figure}[ht]  
\plotone{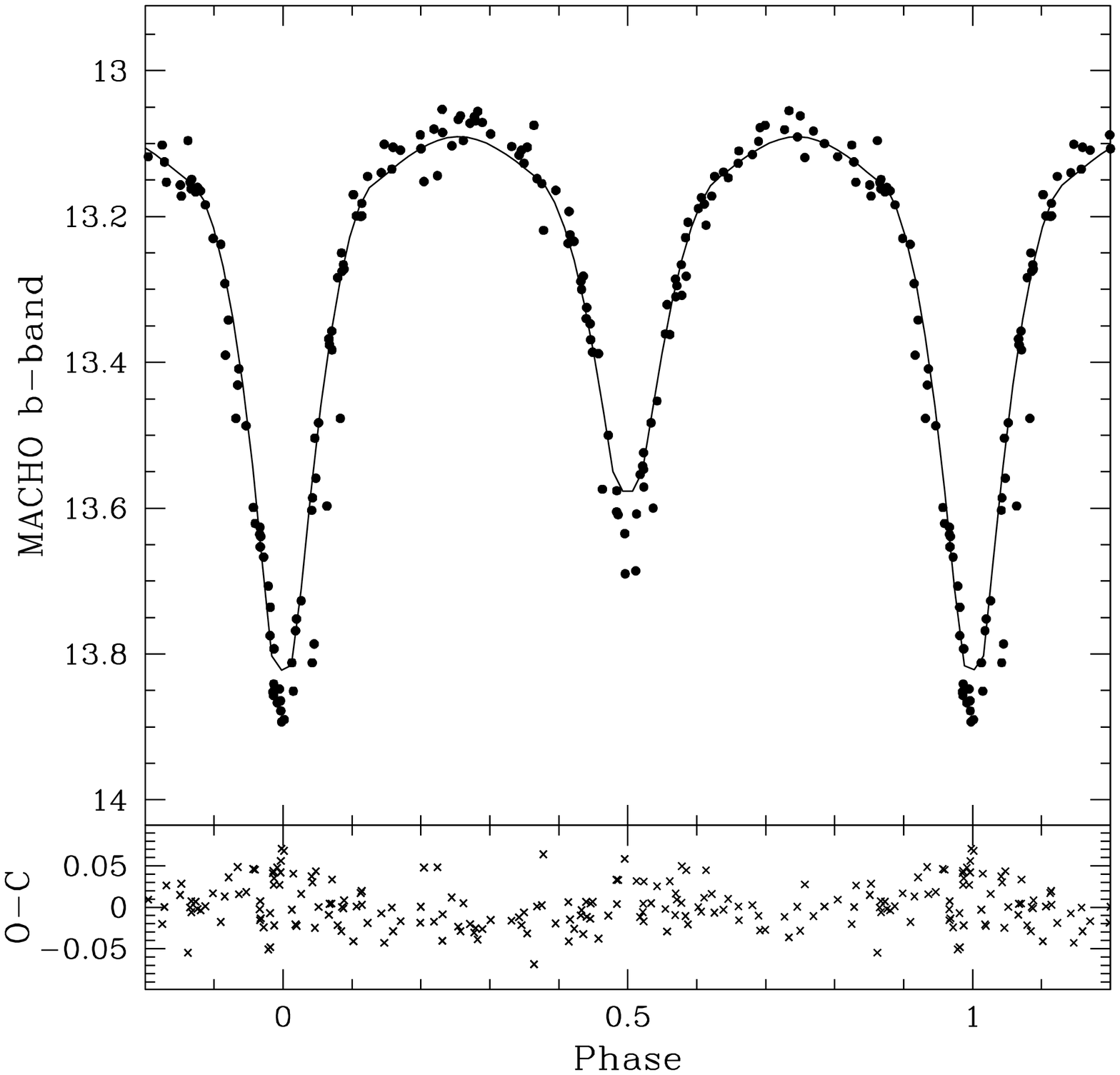}
\caption{Phased MACHO $b$ light curve of LMC-SC1-105 offset to match
OGLE $I-$band photometry; the best fit model from PHOEBE (solid curve)
is overplotted.}
\label{phoebemachobLC}
\end{figure}

The values and errors of the parameters and the stability of the
solution were explored using PHOEBE's scripter. The Wilson-Devinney
differential corrections minimization was run 1000 times, each time
updating the input parameters to the values determined in the previous
iteration. The final values for the parameters were determined by
calculating the mean and standard deviation of the resulting values from
each iteration; the resolution step was adopted as the error for the
period. The errors are similar to the conventional errors from
Wilson-Devinney. A semi-detached configuration with the secondary
filling its Roche lobe yielded a consistent
model. Figures~\ref{lmc105i}, \ref{phoebemachorLC}
and~\ref{phoebemachobLC} present the phased OGLE and MACHO light curves
and best fit model from PHOEBE. The more accurate and better sampled
OGLE light curve exhibits a depression before the primary eclipse, which
is often seen in Algol-type binaries \citep[e.g., see][]{Hilditch05} and
is attributed to a mass-transfer stream. The \citet{O'Connell51} effect
explains the residuals near phase 0.25, since the first quadrature is
brighter than the second by 0.01 mag. The earlier spectral type derived
at this phase for the primary correlates with this extra flux. All bands
display deeper eclipses than predicted by the model. This could be due
to an inaccurate conversion of the differential flux light curves to
magnitudes caused by blending \citep[see][]{Zebrun01}, however it is
also seen in the MACHO light curves, obtained with profile fitting
photometry. The only way to obtain a model with flat residuals (without
changing the mass ratio) would be to additionally model cool spots. This
was not pursued, as adding free parameters would improve the fit, but
not significantly change the masses and radii.

\begin{figure}[ht]  
\plotone{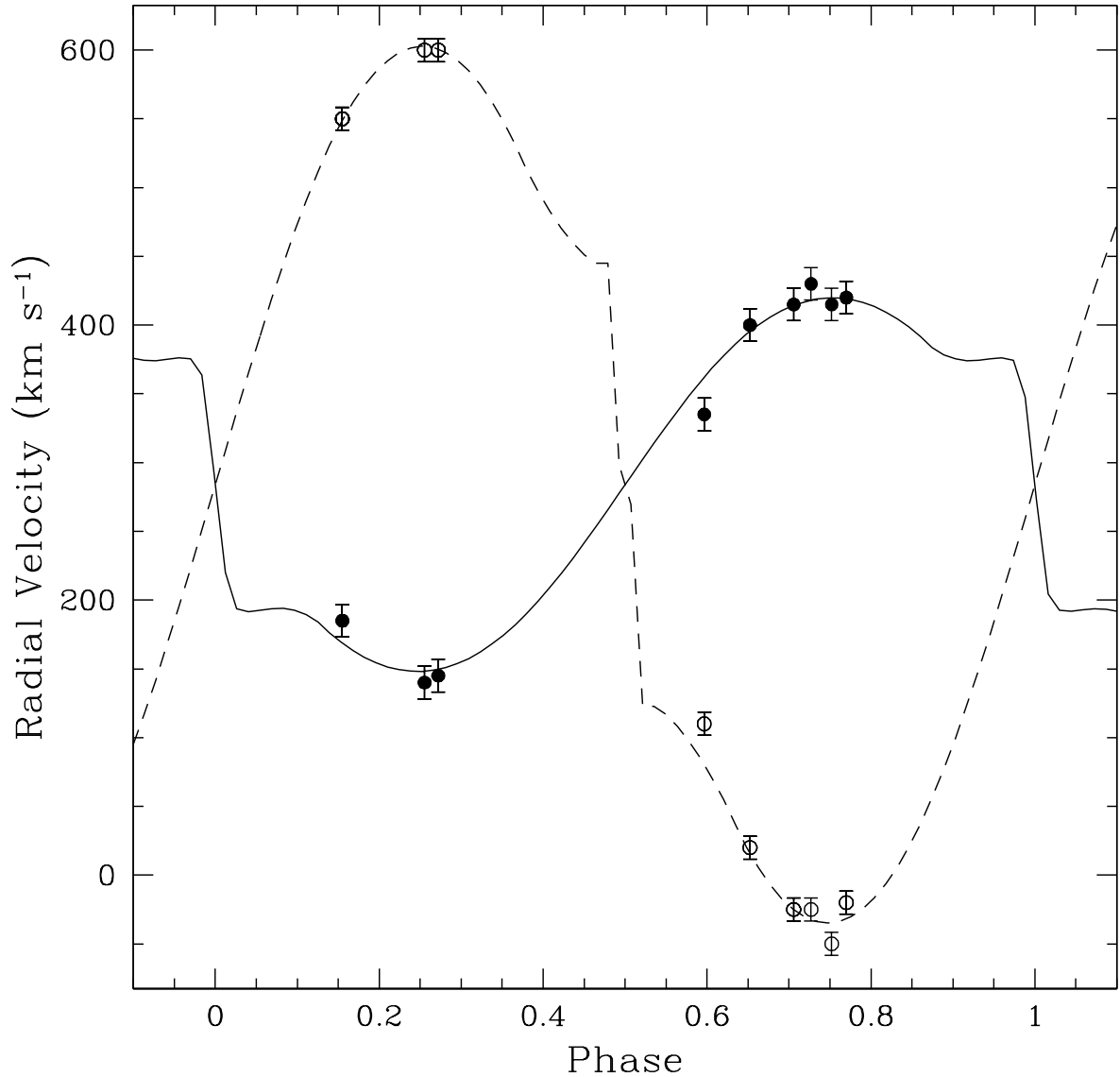}
\caption{Radial velocity curve for LMC-SC1-105. The TODCOR measurements
are shown as filled circles for the primary and open circles for the
secondary; overplotted is the best fit model from PHOEBE, denoted by a
solid line for the primary and a dashed line for the secondary. Error
bars correspond to the rms of the orbital fit: 11.8 $\rm km\; s^{-1}$
for the primary and 8.4 $\rm km\; s^{-1}$ for the secondary.}
\label{lmc105rv}
\end{figure}

\begin{figure}[ht]  
\plotone{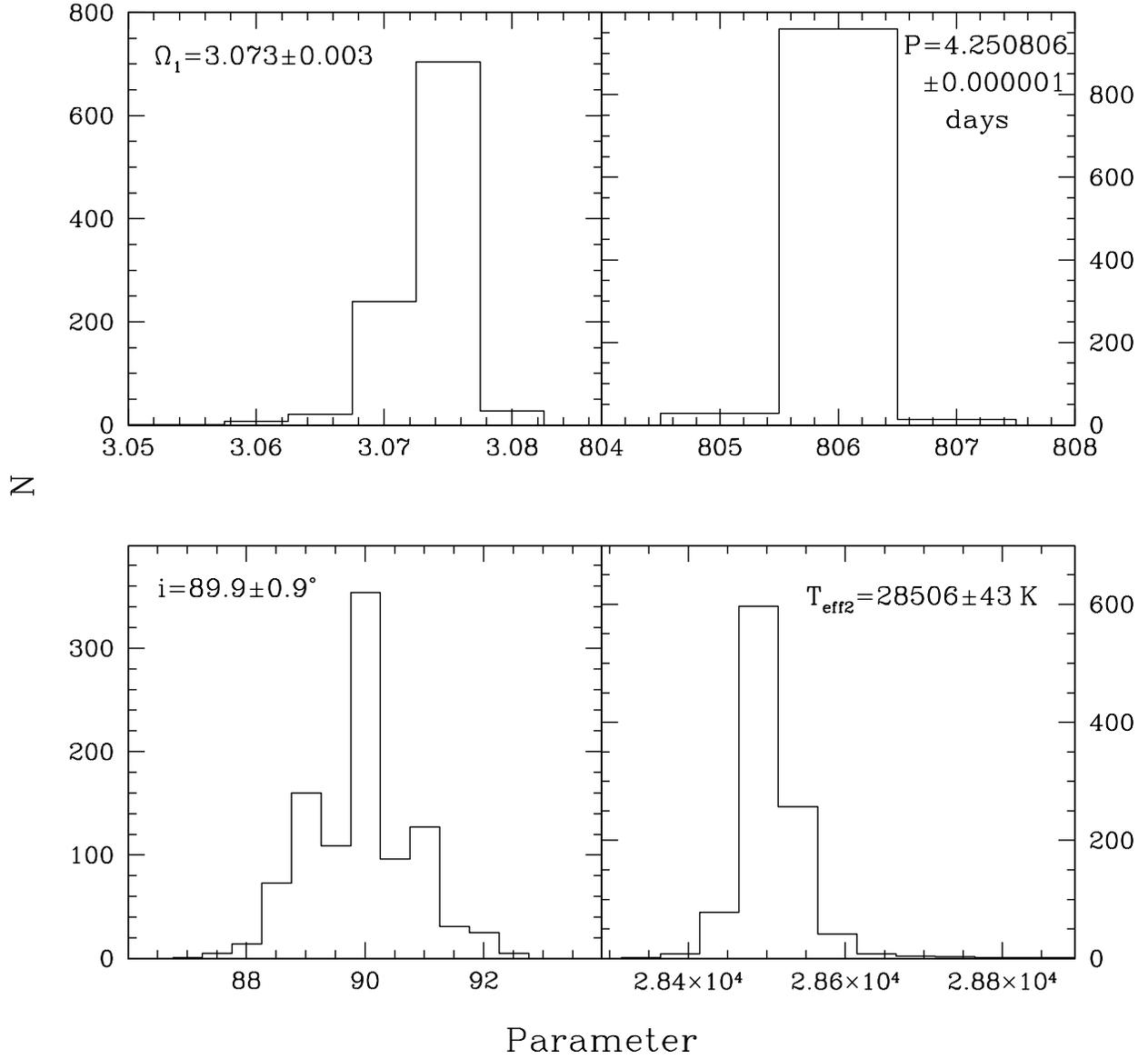}
\caption{Histogram of the results of the heuristic scan for the 4
parameters fit for in the semi-detached mode with PHOEBE. The average
and standard deviation of each parameter is labeled. The axis for $P$
has been scaled for display purposes. The final value for T$_{\rm eff2}$
was adopted from the spectral type calibration.}
\label{lmc105phoebe}
\end{figure}

Figure~\ref{lmc105rv} shows the radial velocity curve and best fit model
from PHOEBE, while Figure~\ref{lmc105phoebe} presents histograms of the
PHOEBE scan results and the final parameters, which are also listed in
Table~\ref{wd105}. The ephemeris is:

\begin{equation}
\rm T_0(HJD)=2450451.90113 + 4.250806(1) \times E,
\end{equation}

\noindent refining the recent period determinations of \citet{Derekas07}
and \citet{Faccioli07} that are based on the MACHO data. The period was
also calculated using the analysis of variance algorithm of
\citet{Schwarzenberg89} on the OGLE light curve and was found to be
4.25083 days. The uncertainties for the parameters are similar to the
formal errors from PHOEBE. T$_{\rm eff2}$ was left as an adjustable
parameter in PHOEBE as a proxy for the flux ratio, however, its value
and uncertainty are quoted from the spectral type calibration described
in \S\ref{section:sse}. The light (or luminosity) ratio is $L_{2}/L_{1}=
0.46\pm0.02$ in $I$, $0.45\pm0.04$ in $R$ and $0.45\pm0.05$ in $V$ and
was fixed to 0.45 in the TODCOR analysis.  These formal errors were
computed by iterating PHOEBE one more time with the primary light levels
as additional adjustable parameters, as described by \citet{Prsa05}. The
flux ratio is $F_{2}/F_{1}=0.74\pm0.03$, in agreement with the value
($0.739 \pm 0.004$) found for LMC-SC1-105 by \citet{Mazeh06} in their
automatic analysis of OGLE eclipsing binaries.

The physical parameters for the system are presented in
Table~\ref{physpar105}, with the final values for the masses and radii
of $\rm M_{1}=30.9\pm1.0\;\msun$ and $\rm R_{1}=15.1\pm0.2\;\rsun$ for
the primary, and $\rm M_{2}=13.0\pm0.7\;\msun$ and $\rm
R_{2}=11.9\pm0.2\;\rsun$ for the secondary. The $\log(g)$ value for the
secondary is lower than that for the primary, in support of the star
being evolved with a luminosity class between III-V. Given the physical
parameters, the synchronous rotational velocities of the stars are found
to be 180 km~s$^{-1}$ and 143 km~s$^{-1}$. Note, the spectra were
rotationally broadened to essentially identical values in
\S\ref{section:sse}. The measured masses and radii of the components
have been determined to $\sim5\%$ and $\sim2\%$ accuracy,
respectively. As discussed in \S1, only 14 accurate measurements
currently exist for very massive stars, with this work contributing
towards increasing this sample to 15 (see Table~\ref{massradiusdata}).

\section{Evolutionary Status}

\begin{figure}[ht]  
\plotone{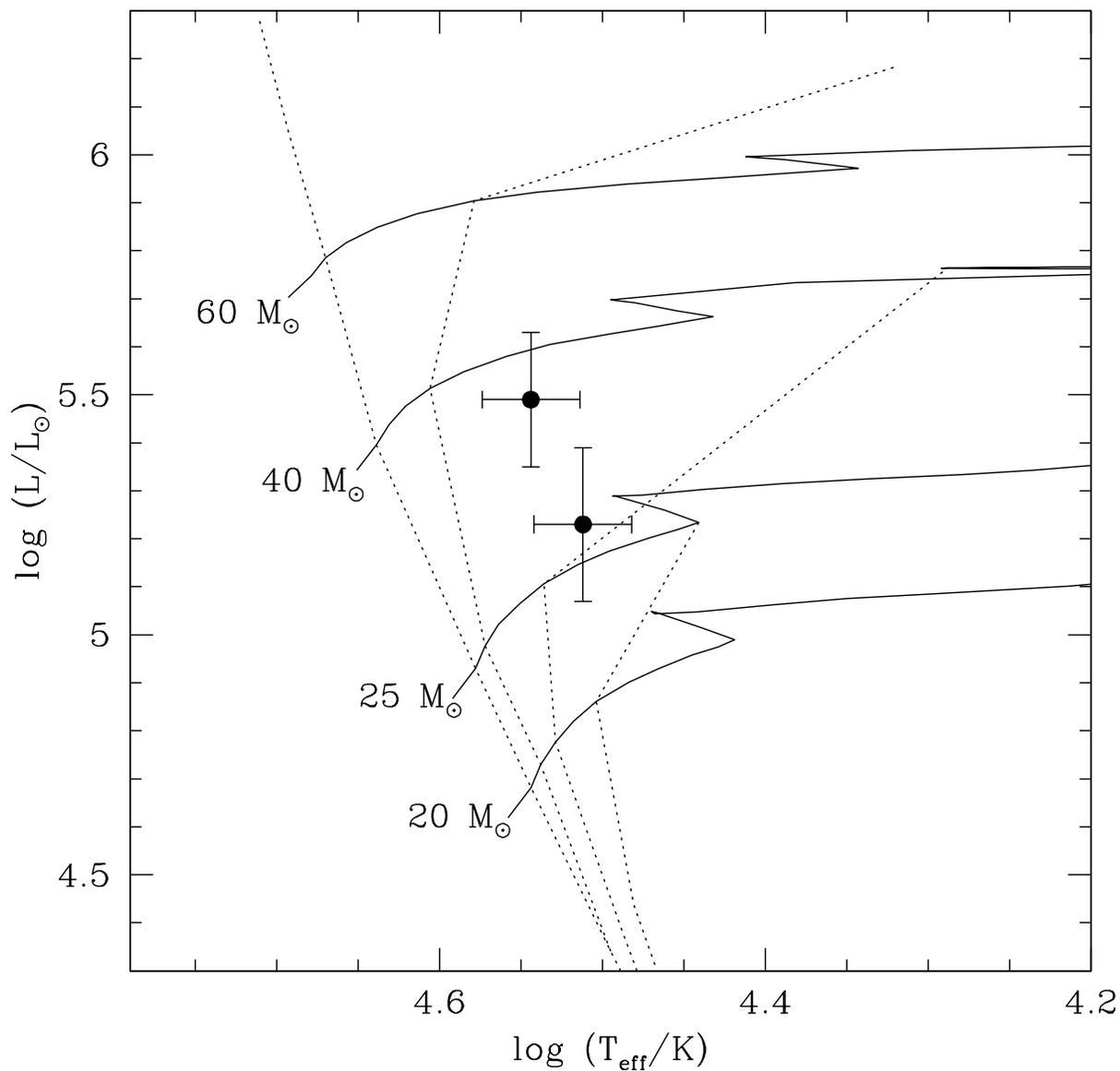}
\caption{Comparison of the parameters of LMC-SC1-105 with evolutionary
tracks (solid lines) and isochrones for single stars at Z=0.008
\citep{Schaerer93}. The dotted lines, from left to right, correspond to
1, 3, 5, 7 and 10 Myr isochrones. Both components are overluminous for
their masses. The cooler, lower mass secondary appears older than the
primary, indicating that the system has undergone mass-transfer.}
\label{lumteff105}
\end{figure}

\begin{figure}[ht]  
\plotone{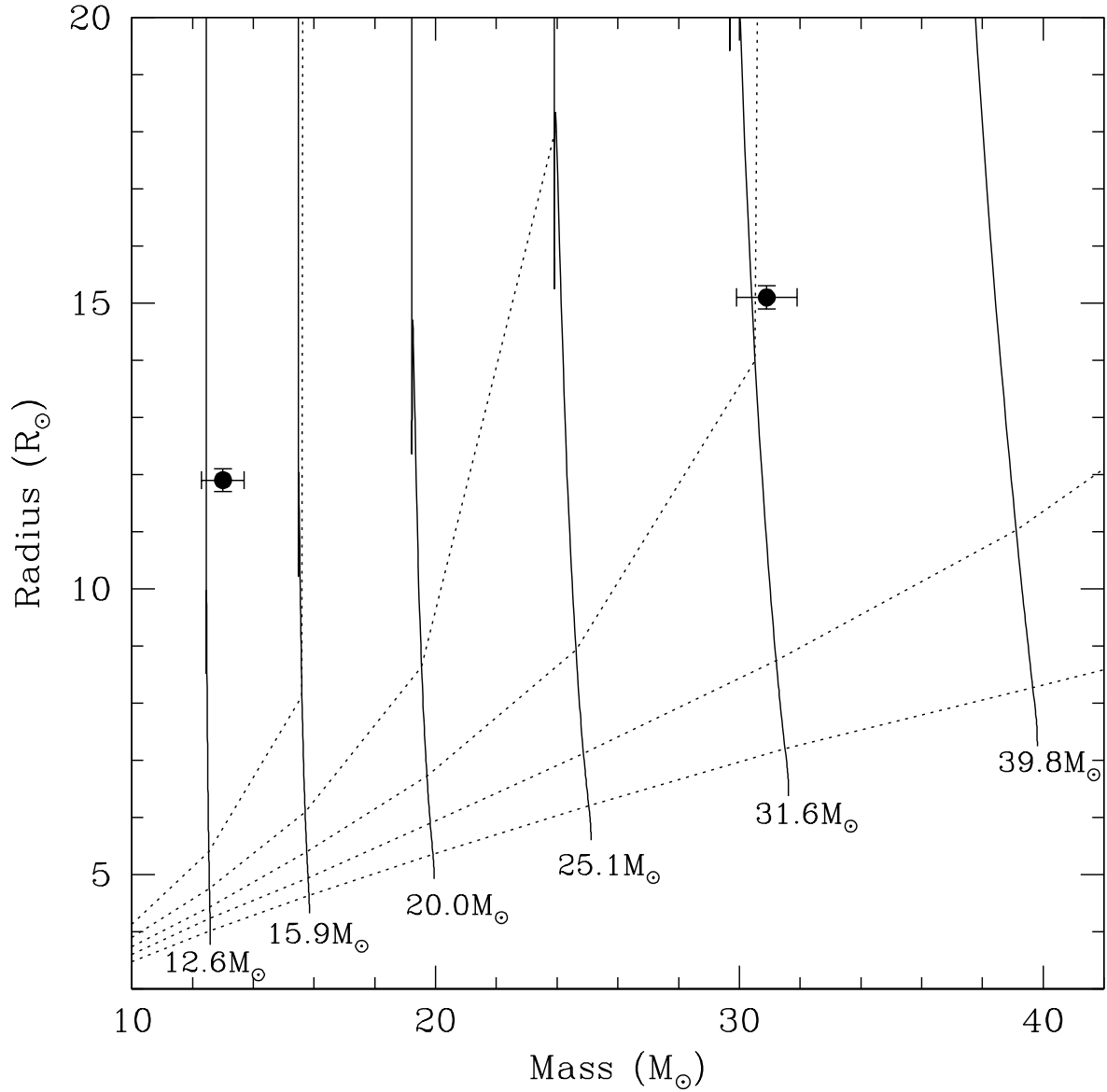}
\caption{Comparison of the parameters of LMC-SC1-105 with evolutionary
tracks and isochrones for single stars at Z=0.007 \citep{Claret06}. The
dotted lines, from the bottom up, correspond to 1, 3, 5, 7 and 10 Myr
isochrones. Single star isochrones are not compatible with the measured
parameters for the system, which has undergone mass-transfer.}
\label{radmass105}
\end{figure}

The semi-detached configuration of LMC-SC1-105 with the less massive
star filling its Roche lobe, along with the main sequence classification
of the primary and possible (sub)giant classification of the secondary,
point to the system being in a slow-mass-transfer stage of case A binary
evolution. The depression in the light curve before primary eclipse
(described in \S\ref{section:lc}) also indicates the presence of a
mass-transfer stream. The emission in the \ion{He}{2} $\lambda$4686 line
and some of the Balmer lines (in particular H$\alpha$) originates near
the secondary and can be explained by gas being transfered onto the
primary. The velocities of these emission lines coincide with the
secondary star but are shifted, indicating an origin near the Lagrangian
L1 point (see Figures~\ref{fitmultioct23} and~\ref{fitmultioct25}). The
evolutionary state of the stars can also be inferred by comparison with
theoretical stellar evolution models. Figures~\ref{lumteff105} and
\ref{radmass105} compare the parameters of the binary with the widely
used theoretical stellar evolutionary models by \citet{Schaerer93} for
single stars and the newer models by \citet{Claret06}, respectively. The
plots illustrate that these post-mass transfer stars are oversized and
overluminous for their masses. According to the isochrones of
\citet{Claret06} in a mass-radius diagram for single stars, the primary
of LMC-SC1-105 has an age $\sim5$ Myr, while the age of the secondary is
$>10$ Myr (see Figure~\ref{radmass105}), which is not in agreement with
them being coeval. This apparent age discrepancy can be explained by
mass-transfer. The ages of the stars according to the evolutionary
models of \citet{Schaerer93} in a luminosity-temperature diagram are
$\sim4$ Myr for the primary and $\sim5$ Myr for the secondary.

\citet{Massey00} have estimated the age and mass of LMC-SC1-105 (or
LH~81-72, W28-22), which is located in the LH~81 association. They find
a range of log(age[yr]) between 5.55-6.77 for the highest mass unevolved
stars and a turnoff mass around $70\,\msun$. Figure~\ref{radmass105}
indicates an age of $\sim5$ Myr or log(age[yr])$\sim6.7$ for the
primary, in agreement with the value of 6.61 derived by \citet{Massey00}
for LMC-SC1-105. They also find a range of reddenings $E(B-V)=0.13-0.23$
mag for LH~81 and a mean value of 0.15 mag. Late O-stars have
$B-V\sim-0.3$, therefore the measured color of LMC-SC1-105
($B-V=-0.16$~mag) implies a reddening value $\sim$0.14
mag. Interestingly, LH~81 contains several evolved stars: the WC4 star
BR~50, the WN4+OB star BR~53 and the B0I+WN star Sk -$69^\circ$194
\citep{Massey00}. LMC-SC1-105 is found by these authors to have a mass
of $39\,\msun$ assuming it is single. This value is 26\% greater than
the dynamical mass of the primary or 12\% smaller than the total mass of
its components, illustrating the systematic error encountered in the
study of massive stars, which could also affect estimations of the
masses of their parent clusters.

Given that LMC-SC1-105 has exchanged mass, a comparison with models of
close binary evolution is needed to infer the initial masses of the
components, however models at the metallicity of the LMC do not
currently exist. According to the solar metallicity evolutionary models
of \citet{Nelson01}, LMC-SC1-105 could have had the following initial
masses, listed as pairs of primary and secondary masses: (25.1, 22.4),
(39.8, 35.5), (50.0, 44.6) \msun, implying an initial mass ratio of
0.90. Lower masses for the initial primary do not result in the measured
mass ratio. These models assume conservative mass-transfer, however the
sum of predicted inital masses ($>$47\,\msun) is greater than the sum of
the final, measured masses (44.1\,\msun). Furthermore, \cite{deMink07}
have shown that conservative evolution is not a valid assumption, by
comparing 50 SMC eclipsing binaries \citep[from][]{Harries03,
Hilditch05} with their grid of models for a range of mass-transfer
efficiencies. A similar grid at the metallicity of the LMC is therefore
necessary to estimate the initial parameters of the binary.

What will the end product of the evolution of LMC-SC1-105 be?  The
primary has a large fill-out ratio of $F=0.89$, as defined by
\citet{Mochnacki72}, implying that it will quickly fill its Roche lobe
when it leaves the main sequence and the hydrogen shell burning phase
begins. Further modeling of LMC-SC1-105, following \citet{deMink07} or
\citet{Petrovic05} for the mass-transfer efficiency, could indicate if
the secondary will evolve into a Wolf-Rayet star and subsequently
explode as a core-collapse supernova or a common envelope phase will
cause the stars to merge.

\section{Conclusions}

This paper presents accurate (to better than $5\%$) fundamental
parameters of LMC-SC1-105, one of the brightest blue eclipsing binary
stars in the LMC found by the OGLE survey. The aim of this work is
twofold: 1) to demonstrate that targeted surveys of the brightest blue
eclipsing binaries in nearby galaxies do indeed select very massive
stars and 2) to measure accurate parameters for one of these rare
systems. The parameters of LMC-SC1-105 were determined from the light
curves available from the OGLE and MACHO surveys and newly acquired high
resolution spectroscopy that targeted quadrature phases, in part
applying the strategy proposed by \citet{Gonzalez05} to constrain the
radial velocity curve with a small number of spectra. The system was
found to contain a very massive main sequence primary
($30.9\pm1.0\,\msun$) and a possibly evolved Roche lobe-filling
secondary. The spectra display the Struve-Sahade effect, which is
present in all the \ion{He}{1} lines, causing the spectral
classification to change with phase, and could be related to the mass
transfer occurring in the system. LMC-SC1-105 could further be used as a
distance indicator to the LMC. However, in addition to accurate radii,
accurate flux (i.e. effective temperatures) and extinction estimates are
necessary for accurate distances. Eclipsing binaries have been used to
derive accurate and independent distances to the LMC
\citep[e.g.][]{Guinan98, Fitzpatrick03}, the Small Magellanic Cloud
\citep{Harries03,Hilditch05}, M31 \citep{Ribas05} and most recently to
M33 \citep{Bonanos06}.

The accurate parameters determined herein for LMC-SC1-105 contribute
valuable data on very massive stars, increasing the current sample of 14
very massive stars with accurate parameters to 15, which despite their
importance remain poorly studied. Such data serve as an external check
to resolve the ``mass discrepancy'' problem, as \citet{Burkholder97}
have shown, and to constrain stellar atmosphere, evolution and formation
models. Further systematic studies of massive binaries in nearby
galaxies are needed to extend the sample of 50 SMC eclipsing binaries
\citep{Harries03, Hilditch05} to higher masses and metallicities and
populate the sparsely sampled parameter space (mass, metallicity,
evolutionary state) with accurate measurements of their masses and
radii. The method of targeting very massive stars in bright blue
eclipsing binaries can therefore be employed towards this goal.

\acknowledgments{I am very grateful to Kris Stanek for motivating me to
undertake this project. I thank the referee, Ian Howarth, for a careful
reading of the manuscript, John Debes for obtaining the Magellan spectra
in October 2007, Igor Soszynski for making available the OGLE-III light
curve prior to publication, Andrej Pr{\v s}a and Phil Massey for
valuable comments that improved the manuscript, Guillermo Torres for his
program implementing TODCOR, Nolan Walborn for advice on the spectral
classification and Ignasi Ribas for his script to merge echelle
orders. This paper utilizes public domain data obtained by the MACHO
Project, jointly funded by the US Department of Energy through the
University of California, Lawrence Livermore National Laboratory under
contract No. W-7405-Eng-48, by the National Science Foundation through
the Center for Particle Astrophysics of the University of California
under cooperative agreement AST-8809616, and by the Mount Stromlo and
Siding Spring Observatory, part of the Australian National
University. AZB acknowledges research and travel support from the
Carnegie Institution of Washington through a Vera Rubin Fellowship.}

{\it Facilities:} \facility{Magellan:Clay (MIKE)}, \facility{Du Pont (Echelle)}

\clearpage
\input{tab1.tex}
\clearpage
\input{tab2.tex}
\input{tab3.tex}
\input{tab4.tex}
\input{tab5.tex}
\end{document}

%% file: tab1.tex
\begin{deluxetable}{llllll}
\tabletypesize{\footnotesize}
\tablewidth{0pc}
\tablecaption{\sc Mass-Radius Data for Very Massive Stars in Eclipsing Binaries}
\tablehead{
\colhead{Eclipsing Binary} & \colhead{Mass} &\colhead{$\sigma_M$} &
\colhead{Radius} & \colhead{$\sigma_R$}& \colhead{Reference}\\
\colhead{Name$^a$} & \colhead{(\msun)} &\colhead{(\msun)} &
\colhead{(\rsun)} & \colhead{(\rsun)}& \colhead{}}
\startdata
FO15 A  & 	30.0  &  1.0	  &   16.0  &  1.0  & \citet{Niemela06} \\
LS1135 A  &    30.0  & 	  1.0  & 12.0  & 1.0  & \citet{Fernandez06} \\
V1007 Sco B &   30.1 &  0.4 &	  15.3 &  0.5 &  \citet{Mayer08}\\	
V1182 Aql A &   31.0 &  0.6 &	  9.00 &  0.18 & \citet{Mayer05} \\
WR~20a B & 82.0 &  5.0 &      19.3 &  0.3 & \citet{Rauw04,Bonanos04} \\	    
WR~20a A & 83.0 &  5.0 &      19.3 &  0.3 & \citet{Rauw04,Bonanos04} \\	    
\hline
LMC-SC1-105 A &         30.9 &  1.0 &   15.1 &  0.2 & This Work \\
LMC Sk-67$^\circ$105 B&	31.4 & 0.7&	13.8 & 0.4  & \citet{Ostrov03} \\
LMC R136-42 B &	32.6 & 0.1&	 6.7 & 0.7  & \citet{Massey02} \\       
LMC HV 2241 A & 	36.2 &  0.7 &	14.9 &  0.4 & \citet{Ostrov01b} \\ 
LMC R136-42 A &	40.3 & 0.1&	 7.4 & 0.8   & \citet{Massey02} \\
LMC MACHO 053441.3 A&	41.2 & 1.2&	9.56 & 0.02 & \citet{Ostrov01c} \\
LMC Sk-67$^\circ$105 A&	48.3 & 0.7&	16.9 & 0.4  & \citet{Ostrov03} \\
LMC R136-38 A&	56.9 & 0.6&	 9.3 &  1.0 & \citet{Massey02} \\
\hline
M33 X-7 & 70.0 &  6.9 &    19.6 &  0.9 & \citet{Orosz07} \\	
\enddata                         
\label{massradiusdata}
\tablenotetext{a}{A and B denote the primary and secondary components of
the EB.}
\end{deluxetable} 

%% file: tab2.tex
\begin{deluxetable}{lllc}
\tabletypesize{\footnotesize}
\tablewidth{0pc}
\tablecaption{\sc Log of Spectroscopic Observations}
\tablehead{
\colhead{UT} & \colhead{Telescope/} & \colhead{Exp. time} & \colhead{S/N at}\\
\colhead{Date} & \colhead{Instrument} & \colhead{(sec)}& \colhead{4500\AA}}
\startdata
20051202 & DuPont/Echelle    & 2$\times$1200  & 45 \\
20051204 & DuPont/Echelle    & 2$\times$1200  & 60 \\
20051206 & DuPont/Echelle    & 2$\times$1200  & 50 \\
20060126 & Clay/MIKE         & 1$\times$200   & 30 \\
20060131 & DuPont/Echelle    & 1$\times$1200  & 40 \\
20060202 & DuPont/Echelle    & 1$\times$1200  & 35 \\
20060204 & DuPont/Echelle    & 1$\times$1200  & 40 \\
20071024 & Clay/MIKE         & 1$\times$600   & 90 \\
20071026 & Clay/MIKE         & 1$\times$600   & 80 \\
\enddata                         
\label{speclog}
\end{deluxetable} 

%% file: tab3.tex
\begin{deluxetable}{cccrrr}
\tabletypesize{\footnotesize}
\tablewidth{0pc}
\tablecaption{\sc Radial Velocity Measurements}
\tablehead{\colhead{HJD} & \colhead{Phase} & \colhead{RV$_{1}$} &
\colhead{$(O-C)_1$} & \colhead{RV$_{2}$} & \colhead{$(O-C)_2$} \\
\colhead{$-$2450000} & \colhead{$\phi$} & \colhead{(km~s$^{-1}$)} &
\colhead{(km~s$^{-1}$)} & \colhead{(km~s$^{-1}$)} &
\colhead{(km~s$^{-1}$)}}
\startdata
3706.76777 &  0.71 & 415 & 1  &  $-$25  & 0  \\
3708.67764 &  0.16 & 185 & 16 &    550  & 1  \\
3710.79198 &  0.65 & 400 & 5  &     20  & 4  \\
3761.56453 &  0.60 & 335 &$-$27 &  110  & 29 \\
3766.55222 &  0.77 & 420 & 1  &  $-$20  & 12 \\
3768.61519 &  0.26 & 140 & $-$8 &  600  & $-$3 \\
3770.62013 &  0.73 & 430 & 12 &  $-$25  & 8  \\
4397.80634 &  0.27 & 145 & $-$5 &  600  & $-$1  \\
4399.84620 &  0.75 & 415 & $-$5 &$-$50  & $-$15\\
\enddata                         
\label{rv}
\end{deluxetable} 

%% file: tab4.tex
\begin{deluxetable}{lc}
\tabletypesize{\footnotesize}
\tablewidth{0pc}
\tablecaption{\sc Results From Light and Radial Velocity Curve Analysis}
\tablehead{
\colhead{Parameter} & \colhead{Value}} 
\startdata
Period, P & 4.250806 $\pm$ 0.000001 days \\
Inclination, $i$        & 89.9 $\pm$ 0.9$^\circ$\\
Surface potential, $\Omega_{1}$ & 3.073 $\pm$ 0.003  \\
Light ratio in $I$, $L_{2}/L_{1}$ & 0.45 $\pm$ 0.02 \\
Mass ratio, $q$ & 0.42 $\pm$ 0.02 \\
Systemic velocity, $\gamma$ & $284\pm3\; \rm km\; s^{-1}$ \\
Semi-major axis, $a$ & $38.9 \pm0.5\; \rsun $  \\
Semi-amplitude, $K_{1}$ & $137\pm4\; \rm km\; s^{-1}$ \\
Semi-amplitude, $K_{2}$ & $326\pm3\; \rm km\; s^{-1}$ \\
Fill-out ratio, F$_1$ & 0.885 $\pm$ 0.003 \\
Radius, $\rm r_{1,pole}$ & 0.373   $\pm$ 0.001 \\     
............ $\rm r_{1,point}$& 0.416   $\pm$ 0.001  \\
............ $\rm r_{1,side}$ & 0.389   $\pm$ 0.001  \\
............ $\rm r_{1,back}$ & 0.402   $\pm$ 0.001  \\
............ $\rm r_{1}$$^{a}$ &  0.389 $\pm$ 0.001 \\

\enddata
\tablenotetext{a}{Volume radius.}
\label{wd105}
\end{deluxetable}

%% file: tab5.tex
\begin{deluxetable}{lcc}
\tabletypesize{\footnotesize}
\tablewidth{0pc}
\tablecaption{\sc Physical Parameters}
\tablehead{
\colhead{Parameter} & \colhead{Primary} & \colhead{Secondary}} 
\startdata
Mass ($\msun$) &  $30.9 \pm1.0$ & $13.0\pm0.7$ \\
Radius ($\rsun$) & $15.1 \pm 0.2$ &  $11.9\pm 0.2$ \\
$\log (g)$ (cgs) &  $3.57 \pm 0.13$ & $3.40\pm 0.20$ \\
$\rm T_{eff}$ (K) & $35000 \pm 2500$ & $32500\pm 2500$ \\
$\log (L/\lsun)$ & $5.49\pm0.14$ & $5.23\pm0.16$ \\ 
\enddata
\label{physpar105}
\end{deluxetable}

%% file: ms.bbl
\clearpage
\begin{thebibliography}{85}
\expandafter\ifx\csname natexlab\endcsname\relax\def\natexlab#1{#1}\fi

\bibitem[{{Andersen}(1991)}]{Andersen91}
{Andersen}, J. 1991, \aapr, 3, 91

\bibitem[{{Bagnuolo} {et~al.}(1999){Bagnuolo}, {Gies}, {Riddle}, \&
  {Penny}}]{Bagnuolo99}
{Bagnuolo}, Jr., W.~G., {Gies}, D.~R., {Riddle}, R., \& {Penny}, L.~R. 1999,
  \apj, 527, 353

\bibitem[{{Bally} \& {Zinnecker}(2005)}]{Bally05}
{Bally}, J. \& {Zinnecker}, H. 2005, \aj, 129, 2281

\bibitem[{{Bernstein} {et~al.}(2003)}]{Bernstein03}
{Bernstein}, R. {et~al.} 2003, in SPIE, Vol. 4841, pp. 1694-1704, ed. M.~{Iye}
  \& A.~F.~M. {Moorwood}

\bibitem[{{Bonanos}(2007)}]{Bonanos07}
{Bonanos}, A.~Z. 2007, \aj, 133, 2696

\bibitem[{{Bonanos} {et~al.}(2006){Bonanos}, {Stanek}, {Kudritzki},
  {et~al.}}]{Bonanos06}
{Bonanos}, A.~Z., {Stanek}, K.~Z., {Kudritzki}, R.~P., {et~al.} 2006, \apj,
  652, 313

\bibitem[{{Bonanos} {et~al.}(2004){Bonanos}, {Stanek}, {Udalski},
  {et~al.}}]{Bonanos04}
{Bonanos}, A.~Z., {Stanek}, K.~Z., {Udalski}, A., {et~al.} 2004, \apjl, 611,
  L33

\bibitem[{{Bouret} {et~al.}(2003){Bouret}, {Lanz}, {Hillier},
  {et~al.}}]{Bouret03}
{Bouret}, J.-C., {Lanz}, T., {Hillier}, D.~J., {et~al.} 2003, \apj, 595, 1182

\bibitem[{{Bromm} \& {Loeb}(2006)}]{Bromm06}
{Bromm}, V. \& {Loeb}, A. 2006, \apj, 642, 382

\bibitem[{{Burkholder} {et~al.}(1997){Burkholder}, {Massey}, \&
  {Morrell}}]{Burkholder97}
{Burkholder}, V., {Massey}, P., \& {Morrell}, N. 1997, \apj, 490, 328

\bibitem[{{Claret}(2000)}]{Claret00}
{Claret}, A. 2000, \aap, 363, 1081

\bibitem[{{Claret}(2006)}]{Claret06}
---. 2006, \aap, 453, 769

\bibitem[{{de Mink} {et~al.}(2007){de Mink}, {Pols}, \& {Hilditch}}]{deMink07}
{de Mink}, S.~E., {Pols}, O.~R., \& {Hilditch}, R.~W. 2007, \aap, 467, 1181

\bibitem[{{Derekas} {et~al.}(2007){Derekas}, {Kiss}, \& {Bedding}}]{Derekas07}
{Derekas}, A., {Kiss}, L.~L., \& {Bedding}, T.~R. 2007, \apj, 663, 249

\bibitem[{{Elmegreen}(2000)}]{Elmegreen00}
{Elmegreen}, B.~G. 2000, \apj, 539, 342

\bibitem[{{Faccioli} {et~al.}(2007){Faccioli}, {Alcock}, {Cook},
  {et~al.}}]{Faccioli07}
{Faccioli}, L., {Alcock}, C., {Cook}, K., {et~al.} 2007, \aj, 134, 1963

\bibitem[{{Fern{\'a}ndez Laj{\'u}s} \& {Niemela}(2006)}]{Fernandez06}
{Fern{\'a}ndez Laj{\'u}s}, E. \& {Niemela}, V.~S. 2006, \mnras, 367, 1709

\bibitem[{{Figer}(2005)}]{Figer05}
{Figer}, D.~F. 2005, \nat, 434, 192

\bibitem[{{Fitzpatrick} {et~al.}(2003){Fitzpatrick}, {Ribas}, {Guinan},
  {et~al.}}]{Fitzpatrick03}
{Fitzpatrick}, E.~L., {Ribas}, I., {Guinan}, E.~F., {et~al.} 2003, \apj, 587,
  685

\bibitem[{{Fryer} {et~al.}(2007){Fryer}, {Mazzali}, {Prochaska},
  {et~al.}}]{Fryer07}
{Fryer}, C.~L., {Mazzali}, P.~A., {Prochaska}, J., {et~al.} 2007, \pasp, 119,
  1211

\bibitem[{{Gies}(2003)}]{Gies03}
{Gies}, D.~R. 2003, in IAU Symposium, Vol. 212, A Massive Star Odyssey: From
  Main Sequence to Supernova, ed. K.~{van der Hucht}, A.~{Herrero}, \&
  C.~{Esteban}, 91--+

\bibitem[{{Gonz{\'a}lez} {et~al.}(2005){Gonz{\'a}lez}, {Ostrov}, {Morrell}, \&
  {Minniti}}]{Gonzalez05}
{Gonz{\'a}lez}, J.~F., {Ostrov}, P., {Morrell}, N., \& {Minniti}, D. 2005,
  \apj, 624, 946

\bibitem[{{Guinan} {et~al.}(1998){Guinan}, {Fitzpatrick}, {Dewarf},
  {et~al.}}]{Guinan98}
{Guinan}, E.~F., {Fitzpatrick}, E.~L., {Dewarf}, L.~E., {et~al.} 1998, \apjl,
  509, L21

\bibitem[{{Hadrava}(1995)}]{Hadrava95}
{Hadrava}, P. 1995, \aaps, 114, 393

\bibitem[{{Harries} {et~al.}(2003){Harries}, {Hilditch}, \&
  {Howarth}}]{Harries03}
{Harries}, T.~J., {Hilditch}, R.~W., \& {Howarth}, I.~D. 2003, \mnras, 339, 157

\bibitem[{{Herrero}(2007)}]{Herrero07}
{Herrero}, A. 2007, RevMexAA in press (arxiv:0704.3528)

\bibitem[{{Hilditch}(2001)}]{Hilditch01}
{Hilditch}, R.~W. 2001, {An Introduction to Close Binary Stars} (An
  Introduction to Close Binary Stars, by R.W.~Hilditch.~ Cambridge University
  Press, 2001, 392 pp.)

\bibitem[{{Hilditch} {et~al.}(2005){Hilditch}, {Howarth}, \&
  {Harries}}]{Hilditch05}
{Hilditch}, R.~W., {Howarth}, I.~D., \& {Harries}, T.~J. 2005, \mnras, 357, 304

\bibitem[{{Hillier} \& {Miller}(1998)}]{Hillier98}
{Hillier}, D.~J. \& {Miller}, D.~L. 1998, \apj, 496, 407

\bibitem[{{Howarth} {et~al.}(1997){Howarth}, {Siebert}, {Hussain},
  {et~al.}}]{Howarth97}
{Howarth}, I.~D., {Siebert}, K.~W., {Hussain}, G.~A.~J., {et~al.} 1997, \mnras,
  284, 265

\bibitem[{{Kelson}(2003)}]{Kelson03}
{Kelson}, D.~D. 2003, \pasp, 115, 688

\bibitem[{{Kelson} {et~al.}(2000)}]{Kelson00}
{Kelson}, D.~D. {et~al.} 2000, \apj, 531, 159

\bibitem[{{Krumholz} \& {Thompson}(2007)}]{Krumholz07}
{Krumholz}, M.~R. \& {Thompson}, T.~A. 2007, \apj, 661, 1034

\bibitem[{{Lanz} \& {Hubeny}(2003)}]{Lanz03}
{Lanz}, T. \& {Hubeny}, I. 2003, \apjs, 146, 417

\bibitem[{{Linder} {et~al.}(2007){Linder}, {Rauw}, {Sana}, {et~al.}}]{Linder07}
{Linder}, N., {Rauw}, G., {Sana}, H., {et~al.} 2007, \aap, 474, 193

\bibitem[{{Lucy}(2006)}]{Lucy06}
{Lucy}, L.~B. 2006, \aap, 457, 629

\bibitem[{{Ma{\'{\i}}z Apell{\'a}niz} {et~al.}(2007){Ma{\'{\i}}z
  Apell{\'a}niz}, {Walborn}, {Morrell}, {et~al.}}]{Maiz-Apellaniz07}
{Ma{\'{\i}}z Apell{\'a}niz}, J., {Walborn}, N.~R., {Morrell}, N.~I., {et~al.}
  2007, \apj, 660, 1480

\bibitem[{{Martins} {et~al.}(2005){Martins}, {Schaerer}, \&
  {Hillier}}]{Martins05}
{Martins}, F., {Schaerer}, D., \& {Hillier}, D.~J. 2005, \aap, 436, 1049

\bibitem[{{Massey}(2003)}]{Massey03}
{Massey}, P. 2003, \araa, 41, 15

\bibitem[{{Massey} {et~al.}(2002){Massey}, {Penny}, \& {Vukovich}}]{Massey02}
{Massey}, P., {Penny}, L.~R., \& {Vukovich}, J. 2002, \apj, 565, 982

\bibitem[{{Massey} {et~al.}(2005){Massey}, {Puls}, {Pauldrach},
  {et~al.}}]{Massey05}
{Massey}, P., {Puls}, J., {Pauldrach}, A.~W.~A., {et~al.} 2005, \apj, 627, 477

\bibitem[{{Massey} {et~al.}(2000){Massey}, {Waterhouse}, \&
  {DeGioia-Eastwood}}]{Massey00}
{Massey}, P., {Waterhouse}, E., \& {DeGioia-Eastwood}, K. 2000, \aj, 119, 2214

\bibitem[{{Mayer} {et~al.}(2005){Mayer}, {Drechsel}, \& {Lorenz}}]{Mayer05}
{Mayer}, P., {Drechsel}, H., \& {Lorenz}, R. 2005, \apjs, 161, 171

\bibitem[{{Mayer} {et~al.}(2008){Mayer}, {Harmanec}, {Nesslinger},
  {et~al.}}]{Mayer08}
{Mayer}, P., {Harmanec}, P., {Nesslinger}, S., {et~al.} 2008, \aap, 481, 183

\bibitem[{{Mazeh} {et~al.}(2006){Mazeh}, {Tamuz}, \& {North}}]{Mazeh06}
{Mazeh}, T., {Tamuz}, O., \& {North}, P. 2006, \mnras, 367, 1531

\bibitem[{{Meynet} \& {Maeder}(2003)}]{Meynet03}
{Meynet}, G. \& {Maeder}, A. 2003, \aap, 404, 975

\bibitem[{{Mochnacki} \& {Doughty}(1972)}]{Mochnacki72}
{Mochnacki}, S.~W. \& {Doughty}, N.~A. 1972, \mnras, 156, 51

\bibitem[{{Mokiem} {et~al.}(2007){Mokiem}, {de Koter}, {Evans},
  {et~al.}}]{Mokiem07}
{Mokiem}, M.~R., {de Koter}, A., {Evans}, C.~J., {et~al.} 2007, \aap, 465, 1003

\bibitem[{{Nelson} \& {Eggleton}(2001)}]{Nelson01}
{Nelson}, C.~A. \& {Eggleton}, P.~P. 2001, \apj, 552, 664

\bibitem[{{Niemela} {et~al.}(2006){Niemela}, {Morrell}, {Fern{\'a}ndez
  Laj{\'u}s}, {et~al.}}]{Niemela06}
{Niemela}, V.~S., {Morrell}, N.~I., {Fern{\'a}ndez Laj{\'u}s}, E., {et~al.}
  2006, \mnras, 367, 1450

\bibitem[{{O'Connell}(1951)}]{O'Connell51}
{O'Connell}, D.~J.~K. 1951, Publications of the Riverview College Observatory,
  2, 85

\bibitem[{{Oey} \& {Clarke}(2005)}]{Oey05}
{Oey}, M.~S. \& {Clarke}, C.~J. 2005, \apjl, 620, L43

\bibitem[{{Orosz} {et~al.}(2007){Orosz}, {McClintock}, {Narayan},
  {et~al.}}]{Orosz07}
{Orosz}, J.~A., {McClintock}, J.~E., {Narayan}, R., {et~al.} 2007, \nat, 449,
  872

\bibitem[{{Ostrov}(2001)}]{Ostrov01c}
{Ostrov}, P.~G. 2001, \mnras, 321, L25

\bibitem[{{Ostrov} \& {Lapasset}(2003)}]{Ostrov03}
{Ostrov}, P.~G. \& {Lapasset}, E. 2003, \mnras, 338, 141

\bibitem[{{Ostrov} {et~al.}(2001){Ostrov}, {Morrell}, \&
  {Lapasset}}]{Ostrov01b}
{Ostrov}, P.~G., {Morrell}, N.~I., \& {Lapasset}, E. 2001, \aap, 377, 972

\bibitem[{{Petrovic} {et~al.}(2005){Petrovic}, {Langer}, \& {van der
  Hucht}}]{Petrovic05}
{Petrovic}, J., {Langer}, N., \& {van der Hucht}, K.~A. 2005, \aap, 435, 1013

\bibitem[{{Pinsonneault} \& {Stanek}(2006)}]{Pinsonneault06}
{Pinsonneault}, M.~H. \& {Stanek}, K.~Z. 2006, \apjl, 639, L67

\bibitem[{{Pr{\v s}a} \& {Zwitter}(2005)}]{Prsa05}
{Pr{\v s}a}, A. \& {Zwitter}, T. 2005, \apj, 628, 426

\bibitem[{{Puls} {et~al.}(2005){Puls}, {Urbaneja}, {Venero}, {et~al.}}]{Puls05}
{Puls}, J., {Urbaneja}, M.~A., {Venero}, R., {et~al.} 2005, \aap, 435, 669

\bibitem[{{Pych}(2004)}]{Pych04}
{Pych}, W. 2004, \pasp, 116, 148

\bibitem[{{Rauw} {et~al.}(2004){Rauw}, {De Becker}, {Naze}, {et~al.}}]{Rauw04}
{Rauw}, G., {De Becker}, M., {Naze}, Y., {et~al.} 2004, \aap, 420, L9

\bibitem[{{Repolust} {et~al.}(2004){Repolust}, {Puls}, \&
  {Herrero}}]{Repolust04}
{Repolust}, T., {Puls}, J., \& {Herrero}, A. 2004, \aap, 415, 349

\bibitem[{{Ribas} {et~al.}(2005){Ribas}, {Jordi}, {Vilardell},
  {et~al.}}]{Ribas05}
{Ribas}, I., {Jordi}, C., {Vilardell}, F., {et~al.} 2005, \apjl, 635, L37

\bibitem[{{Santolaya-Rey} {et~al.}(1997){Santolaya-Rey}, {Puls}, \&
  {Herrero}}]{Santolaya97}
{Santolaya-Rey}, A.~E., {Puls}, J., \& {Herrero}, A. 1997, \aap, 323, 488

\bibitem[{{Schaerer} {et~al.}(1993){Schaerer}, {Charbonnel}, {Meynet},
  {et~al.}}]{Schaerer93}
{Schaerer}, D., {Charbonnel}, C., {Meynet}, G., {et~al.} 1993, \aaps, 102, 339

\bibitem[{{Schaller} {et~al.}(1992){Schaller}, {Schaerer}, {Meynet}, \&
  {Maeder}}]{Schaller92}
{Schaller}, G., {Schaerer}, D., {Meynet}, G., \& {Maeder}, A. 1992, \aaps, 96,
  269

\bibitem[{{Schnurr} {et~al.}(2008){Schnurr}, {Casoli}, {Chen{\'e}},
  {et~al.}}]{Schnurr08}
{Schnurr}, O., {Casoli}, J., {Chen{\'e}}, A.-N., {et~al.} 2008, \mnras, 389,
  L38

\bibitem[{{Schwarzenberg-Czerny}(1989)}]{Schwarzenberg89}
{Schwarzenberg-Czerny}, A. 1989, \mnras, 241, 153

\bibitem[{{Simon} \& {Sturm}(1994)}]{Simon94}
{Simon}, K.~P. \& {Sturm}, E. 1994, \aap, 281, 286

\bibitem[{{Soszynski} {et~al.}(2008){Soszynski}, {Poleski}, {Udalski},
  {et~al.}}]{Soszynski08}
{Soszynski}, I., {Poleski}, R., {Udalski}, A., {et~al.} 2008, AcA, submitted
  (astro-ph/0808.2210)

\bibitem[{{Southworth} \& {Clausen}(2007)}]{Southworth07}
{Southworth}, J. \& {Clausen}, J.~V. 2007, \aap, 461, 1077

\bibitem[{{Stickland}(1997)}]{Stickland97}
{Stickland}, D.~J. 1997, The Observatory, 117, 37

\bibitem[{{Tonry} \& {Davis}(1979)}]{Tonry79}
{Tonry}, J. \& {Davis}, M. 1979, \aj, 84, 1511

\bibitem[{{Vitrichenko} {et~al.}(2007){Vitrichenko}, {Nadyozhin}, \&
  {Razinkova}}]{Vitrichenko07}
{Vitrichenko}, E.~A., {Nadyozhin}, D.~K., \& {Razinkova}, T.~L. 2007, Astronomy
  Letters, 33, 251

\bibitem[{{Walborn} \& {Fitzpatrick}(1990)}]{Walborn90}
{Walborn}, N.~R. \& {Fitzpatrick}, E.~L. 1990, \pasp, 102, 379

\bibitem[{{Walborn} {et~al.}(2002){Walborn}, {Howarth}, {Lennon},
  {et~al.}}]{Walborn02}
{Walborn}, N.~R., {Howarth}, I.~D., {Lennon}, D.~J., {et~al.} 2002, \aj, 123,
  2754

\bibitem[{{Wilson}(1979)}]{Wilson79}
{Wilson}, R.~E. 1979, \apj, 234, 1054

\bibitem[{{Wilson}(1990)}]{Wilson90}
---. 1990, \apj, 356, 613

\bibitem[{{Wilson} \& {Devinney}(1971)}]{Wilson71}
{Wilson}, R.~E. \& {Devinney}, E.~J. 1971, \apj, 166, 605

\bibitem[{{Wyrzykowski} {et~al.}(2003){Wyrzykowski}, {Udalski}, {Kubiak},
  {et~al.}}]{Wyrzykowski03}
{Wyrzykowski}, L., {Udalski}, A., {Kubiak}, M., {et~al.} 2003, Acta
  Astronomica, 53, 1

\bibitem[{{Yungelson} {et~al.}(2008){Yungelson}, {van den Heuvel}, {Vink},
  {et~al.}}]{Yungelson08}
{Yungelson}, L.~R., {van den Heuvel}, E.~P.~J., {Vink}, J.~S., {et~al.} 2008,
  \aap, 477, 223

\bibitem[{{Zebrun} {et~al.}(2001){Zebrun}, {Soszynski}, {Wozniak},
  {et~al.}}]{Zebrun01}
{Zebrun}, K., {Soszynski}, I., {Wozniak}, P.~R., {et~al.} 2001, Acta
  Astronomica, 51, 317

\bibitem[{{Zinnecker} \& {Yorke}(2007)}]{Zinnecker07}
{Zinnecker}, H. \& {Yorke}, H.~W. 2007, \araa, 45, 481

\bibitem[{{Zucker} \& {Mazeh}(1994)}]{Zucker94}
{Zucker}, S. \& {Mazeh}, T. 1994, \apj, 420, 806

\end{thebibliography}
